\documentclass[
  pra,
  aps,
  reprint,
  superscriptaddress,
  nofootinbib,
  floatfix
]{revtex4-2}

\usepackage[utf8]{inputenc}
\usepackage[english]{babel}
\usepackage{amsmath}
\usepackage{amssymb}
\usepackage{mathtools}
\usepackage{bm}
\usepackage{graphicx}
\usepackage{xcolor}
\usepackage{booktabs}
\usepackage{hyperref}
\begin{document}

\title{Wave-functional formulation of dissipative CSL models}
\author{Y. M. P. Gomes}
\email{ymuller@cbpf.br}
 \affiliation{Centro Brasileiro de Pesquisas F\'isicas, Rua Dr. Xavier Sigaud 150, Urca, CEP: 22290-180, Rio de Janeiro-RJ, Brazil}








\begin{abstract}
We formulate minimal and dissipative Continuous Spontaneous Localization (CSL) dynamics in the functional Schrödinger representation for a non-relativistic bosonic field. In this framework, the Fock-space state is encoded in a wave functional, and fixed particle-number wave functions are obtained by sector projection. For the minimal CSL coupling to the smeared mass density, this projection gives the standard nonlinear stochastic dynamics in each \(N\)-particle sector, with the collapse operator acting on the total smeared density of the configuration. This makes the amplification mechanism transparent and allows us to discuss sector superpositions, local probability balance, and the status of Bohmian equivariance at the wave-function level. We then consider a dissipative extension in which the collapse operator includes a smeared current contribution. The one-particle sector reproduces the expected dissipative CSL energy balance, while fixed many-body sectors contain additional collective momentum shifts and pair-mixing terms that are not reducible, in general, to independent one-particle contributions. Within a leading compact closure, the collective pair friction produces
a non-extensive stationary mean kinetic energy: in three dimensions and for weak dissipation, $T_N^{\rm comp}\simeq 2T_\beta/N$, whereas the corresponding dilute energy remains extensive.

\end{abstract}

\maketitle
\flushbottom

 
\section{Introduction}
\label{sec:introduction}

The measurement problem arises from the coexistence, in the standard
formulation of quantum mechanics, of deterministic Schrödinger evolution
and non-unitary state reduction during measurement \cite{sakurai}. In the
Copenhagen framework, this prescription is associated with a division
between microscopic quantum systems and macroscopic classical apparatuses,
as well as with Bohr's principle of complementarity \cite{bohr1}. Alternative
formulations, such as de Broglie--Bohm theory, retain definite particle
trajectories and make probability flow and guidance dynamics central to the
description \cite{bohm1,bohm2}. None of these formulations, however, provides
within the standard Schrödinger dynamics an intrinsic mass, size, or
complexity scale at which macroscopic superpositions cease to persist.

Spontaneous-collapse models address this problem by introducing stochastic
and nonlinear modifications of quantum dynamics; see Ref.~\cite{Bassi2013}
for a review. Their main examples include the Ghirardi--Rimini--Weber model
\cite{GRWref}, the Di\'osi--Penrose proposals \cite{DPref1,DPref2}, and
Continuous Spontaneous Localization (CSL) \cite{CSLref,Adlerref}. In CSL,
collapse effects are weak for microscopic systems but are enhanced for
configurations containing many constituents or large masses. This
amplification mechanism makes spatially separated macroscopic
superpositions dynamically unstable while leaving ordinary microscopic
predictions approximately unchanged. The mesoscopic regime is therefore
particularly relevant: experimental bounds depend on the mass, coherence
time, and spatial resolution of the tested superpositions
\cite{NimmrichterHornberger2013}.

At the statistical level, collapse dynamics is commonly represented by a
Lindblad master equation \cite{lindref1,lindref2}. When individual noise
realizations, probability currents, quantum potentials, or trajectory-based
descriptions are relevant, however, a wave-function-level formulation is
more informative. In particular, it allows one to examine how collapse
modifies the local probability balance and the usual Bohmian equivariance
property \cite{holland95}. The functional Schrödinger representation provides
a natural framework for this purpose: the full Fock-space state is encoded
in a wave functional, while ordinary configuration-space wave functions are
obtained by projection onto fixed particle-number sectors. Functional
representations and their Hamilton--Jacobi counterparts have been developed
in several field-theoretic settings \cite{kiefer92,ivanov20,sint94,kieferwipf94}.

In this work, we formulate minimal CSL dynamics for a non-relativistic
bosonic field in the functional Schrödinger representation. The collapse
modification is introduced once at the wave-functional level and then
projected onto fixed-\(N\) sectors. The projected collapse operator acts on
the total smeared density of each configuration, making the amplification
mechanism explicit and distinguishing coherent compact configurations from
dilute ones. The same framework also allows us to discuss superpositions of
different particle-number sectors, the stochastic evolution of their
relative weights, and the replacement of the ordinary continuity equation
by a local stochastic probability balance. Although the total norm is
preserved, the standard Bohmian guidance flow does not, by itself, reproduce
the collapse-induced local redistribution of probability.

We then consider a dissipative extension in which the collapse operator
contains a smeared current contribution. Such extensions have been
introduced to control collapse-induced energy growth and generate
friction-like effects \cite{Smirne2014,Smirne2015,Toros2017}. The
one-particle sector reproduces the expected competition between heating and
dissipation and therefore serves as a consistency check. In fixed
many-particle sectors, however, the projection generates collective momentum
shifts and off-diagonal pair terms that cannot, in general, be reduced to a
sum of independent one-particle contributions. These terms modify both the
Hamilton--Jacobi equation and the energy balance. 


This paper is organized as follows. Section~\ref{sec:bargmann_representation} introduces the functional Schrödinger representation and the projection onto fixed particle-number sectors. Section~\ref{sec:minimal_CSL} develops the minimal CSL dynamics, including amplification, sector superpositions, energy growth, and local probability balance. Section~\ref{sec:dissipative_csl} introduces the dissipative extension, derives the one-particle and fixed-\(N\) equations, and identifies the collective pair contributions. Section~\ref{sec:conclusions} summarizes the results and discusses possible extensions. Throughout the derivations we set \(\hbar=1\), except when restoring physical units in the definition of \(\kappa\) and in the phenomenological estimates.

\section{Bargmann coherent-state functional representation}
\label{sec:bargmann_representation}

We employ the holomorphic Bargmann representation of the bosonic
Fock space. To avoid ambiguities concerning normalization, we work
with unnormalized coherent states,
\begin{equation}
 \|\alpha\rangle
 =
 \exp\left[
 \int d^d x\,
 \alpha(\mathbf{x})\hat a^\dagger(\mathbf{x})
 \right]|0\rangle,
\end{equation}
and their duals,
\begin{equation}
 \langle\bar\alpha\|
 =
 \langle 0|
 \exp\left[
 \int d^d x\,
 \bar\alpha(\mathbf{x})\hat a(\mathbf{x})
 \right].
\end{equation}
They obey
\begin{equation}
 \hat a(\mathbf{x})\|\alpha\rangle
 =
 \alpha(\mathbf{x})\|\alpha\rangle,
 \qquad
 \langle\bar\alpha\|\hat a^\dagger(\mathbf{x})
 =
 \bar\alpha(\mathbf{x})\langle\bar\alpha\|.
\end{equation}

A Fock-space state $|\Psi(t)\rangle$ is represented by the
holomorphic wave functional
\begin{equation}
 \Psi[\bar\alpha,t]
 =
 \langle\bar\alpha\|\Psi(t)\rangle.
\end{equation}
The creation and annihilation operators act as
\begin{equation}
 \hat a^\dagger(\mathbf{x})
 \longrightarrow
 \bar\alpha(\mathbf{x}),
 \qquad
 \hat a(\mathbf{x})
 \longrightarrow
 \frac{\delta}{\delta\bar\alpha(\mathbf{x})}.
\end{equation}
Consequently,
\begin{equation}
 \left[
 \frac{\delta}{\delta\bar\alpha(\mathbf{x})},
 \bar\alpha(\mathbf{z})
 \right]
 =
 \delta^{(d)}(\mathbf{x}-\mathbf{z}),
\end{equation}
which realizes the canonical bosonic algebra.

The normalized coherent states conventionally denoted by
$|\alpha\rangle$ are related to the Bargmann states by
\begin{equation}
 |\alpha\rangle
 =
 \exp\left[
 -\frac{1}{2}
 \int d^d x\,|\alpha(\mathbf{x})|^2
 \right]\|\alpha\rangle.
\end{equation}
Thus the functional used here should not be confused with the
normalized coherent-state amplitude $\langle\alpha|\Psi\rangle$,
which contains an additional Gaussian factor.

The Fock-space inner product can be represented formally as
\begin{equation}
 \langle\Phi|\Psi\rangle
 =
 \int \mathcal D^2\alpha\,
 \exp\left[
 -\int d^d x\,|\alpha(\mathbf{x})|^2
 \right]
 \Phi^*[\bar\alpha]\Psi[\bar\alpha],
\end{equation}
where the functional measure may be defined by first working with
a finite number of modes. In the following, expectation values are
understood as Fock-space expectation values, with the above
expression providing their Bargmann representation.

For the free nonrelativistic Hamiltonian,
\begin{equation}
 \hat H_0
 =
 \int d^d x\,
 \hat a^\dagger(\mathbf{x})
 \left(
 -\frac{\nabla^2}{2m}
 \right)
 \hat a(\mathbf{x}),
\end{equation}
the Schrödinger equation becomes
\begin{equation}
 i\partial_t\Psi[\bar\alpha,t]
 =
 \mathcal H_0 \Psi[\bar\alpha,t],
\end{equation}
with
\begin{equation}
 \mathcal H_0
 =
 \int d^d x\,
 \bar\alpha(\mathbf{x})
 \left(
 -\frac{\nabla^2}{2m}
 \right)
 \frac{\delta}{\delta\bar\alpha(\mathbf{x})}.
\end{equation}

The decomposition of the state into fixed particle-number sectors is
\begin{equation}
 |\Psi(t)\rangle
 =
 \sum_{N=0}^{\infty}|\Psi_N(t)\rangle,
\end{equation}
where
\begin{equation}
 |\Psi_N(t)\rangle
 =
 \frac{1}{\sqrt{N!}}
 \int dX_N\,
 \psi_N(X_N,t)
 \hat a^\dagger(\mathbf{x}_1)\cdots
 \hat a^\dagger(\mathbf{x}_N)|0\rangle.
\end{equation}
Here
\begin{equation}
 X_N=(\mathbf{x}_1,\ldots,\mathbf{x}_N),
 \qquad
 dX_N=\prod_{a=1}^{N}d^d x_a,
\end{equation}
and $\psi_N$ is symmetric under the exchange of any pair of
coordinates. The corresponding Bargmann functional is
\begin{equation}
 \Psi[\bar\alpha,t]
 =
 \sum_{N=0}^{\infty}
 \frac{1}{\sqrt{N!}}
 \int dX_N\,
 \psi_N(X_N,t)
 \prod_{a=1}^{N}\bar\alpha(\mathbf{x}_a).
\end{equation}
The fixed-$N$ wave function is therefore recovered through
\begin{equation}
 {
 \psi_N(X_N,t)
 =
 \frac{1}{\sqrt{N!}}
 \left.
 \frac{\delta^N\Psi[\bar\alpha,t]}
 {\delta\bar\alpha(\mathbf{x}_N)\cdots
  \delta\bar\alpha(\mathbf{x}_1)}
 \right|_{\bar\alpha=0}
 }.
\end{equation}

The Fock-space normalization becomes
\begin{equation}
 \langle\Psi|\Psi\rangle
 =
 \sum_{N=0}^{\infty}P_N(t),
 \qquad
 P_N(t)
 =
 \int dX_N\,|\psi_N(X_N,t)|^2 .
\end{equation}
For a state restricted to a single particle-number sector,
$P_N=1$. For a superposition of sectors, $P_N$ is the probability
weight of the corresponding sector and the individual
$\psi_N$ are not separately normalized.

Since the free Hamiltonian conserves particle number, projection of
the functional Schrödinger equation gives
\begin{equation}
 i\partial_t\psi_N(X_N,t)
 =
 H_N\psi_N(X_N,t),
\end{equation}
where
\begin{equation}
 H_N
 =
 \sum_{a=1}^{N}
 \left(
 -\frac{\nabla_a^2}{2m}
 \right).
\end{equation}
In particular,
\begin{equation}
 i\partial_t\psi_1(\mathbf{x},t)
 =
 -\frac{\nabla^2}{2m}\psi_1(\mathbf{x},t),
\end{equation}
and
\begin{equation}
 i\partial_t\psi_2(\mathbf{x}_1,\mathbf{x}_2,t)
 =
 \left(
 -\frac{\nabla_1^2}{2m}
 -\frac{\nabla_2^2}{2m}
 \right)
 \psi_2(\mathbf{x}_1,\mathbf{x}_2,t).
\end{equation}

The same projection rule applies to number-conserving bilinear
operators. For example,
\begin{equation}
 \hat n(\mathbf{x})
 =
 \hat a^\dagger(\mathbf{x})\hat a(\mathbf{x})
\end{equation}
is represented by
\begin{equation}
 n(\mathbf{x})
 =
 \bar\alpha(\mathbf{x})
 \frac{\delta}{\delta\bar\alpha(\mathbf{x})}.
\end{equation}
For the smeared number-density operator
\begin{equation}
 \hat N_{\mathbf y}
 =
 \int d^d x\,
 L_{\mathbf y}(\mathbf{x})
 \hat a^\dagger(\mathbf{x})\hat a(\mathbf{x}),
\end{equation}
the Bargmann representative is
\begin{equation}
 \mathcal N_{\mathbf y}
 =
 \int d^d x\,
 L_{\mathbf y}(\mathbf{x})
 \bar\alpha(\mathbf{x})
 \frac{\delta}{\delta\bar\alpha(\mathbf{x})}.
 \label{eq:functional_smeared_density}
\end{equation}
Its action in the fixed-$N$ sector is
\begin{equation}
 \hat N_{\mathbf y}
 \longrightarrow
 N_{\mathbf y}^{(N)}(X_N)
 =
 \sum_{a=1}^{N}L_{\mathbf y}(\mathbf{x}_a).
\end{equation}
This identity will be used below to project the CSL dynamics onto
ordinary configuration-space wave functions.
\section{Minimal CSL in the Functional Representation}
\label{sec:minimal_CSL}

We now formulate the standard mass-proportional Continuous
Spontaneous Localization (CSL) dynamics \cite{CSLref,Bassi2013} in
the Bargmann functional representation introduced in the preceding
section. The smeared number-density operator $\hat N_y$, its
functional representative $\mathcal N_y$, and its fixed-sector action
$N_y^{(N)}(X_N)$ have already been defined. For identical bosons of
mass $m$, the corresponding smeared mass-density operator is
\begin{equation}
    \hat M_y=m\hat N_y.
\end{equation}
The spatial correlations are encoded in the localization profile
$L_y(x)\equiv L(x-y)$ through
\begin{equation}
    g(x-z)
    =
    \int d^d y\,L_y(x)L_y(z).
    \label{eq:kernel_definition}
\end{equation}
For a translationally invariant kernel,
$\widetilde g(k)=|\widetilde L(k)|^2$; for a real and even profile one
may choose $\widetilde L(k)=\sqrt{\widetilde g(k)}$.

For a normalized Fock-space state, define
\begin{equation}
    \bar N_y^{\rm tot}(t)
    =
    \langle\Psi_t|\hat N_y|\Psi_t\rangle,
    \qquad
    \lambda
    =
    \frac{\sqrt{\gamma}\,m}{m_0}.
    \label{eq:lambda_definition}
\end{equation}
Here $\lambda$ is the coefficient multiplying the Wiener increment;
with the present conventions, the deterministic CSL contribution is
proportional to $\lambda^2$. The normalized nonlinear It\^o equation
then reads
\begin{align}
    d\Psi[\bar\alpha,t]
    ={}&
    \Bigg[
    -i\mathcal H_0\,dt
    \nonumber\\
    &
    +\lambda\int d^d y\,
    \bigl(\mathcal N_y-\bar N_y^{\rm tot}(t)\bigr)dW_t(y)
    \nonumber\\
    &
    -\frac{\lambda^2}{2}
    \int d^d y\,
    \bigl(\mathcal N_y-\bar N_y^{\rm tot}(t)\bigr)^2dt
    \Bigg]\Psi[\bar\alpha,t].
    \label{eq:CSL_functional}
\end{align}
The stochastic increments satisfy
\begin{equation}
    \mathbb E[dW_t(y)]=0,
    \qquad
    \mathbb E[dW_t(y)dW_t(z)]
    =
    \delta^{(d)}(y-z)\,dt.
    \label{eq:white_noise}
\end{equation}
Thus the noise is spatially white, while all spatial correlations are
carried by the smearing profile. This avoids introducing the same
kernel both in the collapse operator and in the noise correlator.

\subsection{Fixed particle-number sectors}
\label{subsec:minimal_fixed_N}

Consider first a normalized state restricted to the $N$-particle
sector. Using the fixed-sector action $N_y^{(N)}(X_N)$ derived in
Sec.~\ref{sec:bargmann_representation}, we define
\begin{equation}
    \bar N_y(t)
    =
    \int dX_N\,\rho_N(X_N,t)N_y^{(N)}(X_N),\qquad
    \rho_N=|\psi_N|^2.
    \label{eq:Ny_mean_fixed}
\end{equation}
Projection of Eq.~\eqref{eq:CSL_functional} gives
\begin{align}
    d\psi_N(X_N,t)
    &={}
    \Bigg[
    -iH_N\,dt
    \nonumber\\
    &
    +\lambda\int d^d y\,
    \Delta N_y^{(N)}(X_N,t)dW_t(y)
    \nonumber\\
    &
    -\frac{\lambda^2}{2}
    \int d^d y\,
    \bigl[\Delta N_y^{(N)}(X_N,t)\bigr]^2dt
    \Bigg]
    \nonumber\\
    &\times\psi_N(X_N,t),
    \label{eq:CSL_fixed_N}
\end{align}
where
\begin{equation}
    \Delta N_y^{(N)}(X_N,t)
    =
    N_y^{(N)}(X_N)-\bar N_y(t).
    \label{eq:Delta_N_fixed}
\end{equation}
Since $N_y^{(N)}$ is symmetric under permutations of the coordinates,
Eq.~\eqref{eq:CSL_fixed_N} preserves bosonic exchange symmetry.
Moreover, because the collapse operator is number conserving, a state
initially restricted to this sector remains in it.

The collective structure can be displayed through the normalized
empirical density
\begin{equation}
    \nu_{X_N}(x)
    =
    \frac{1}{N}\sum_{a=1}^{N}\delta^{(d)}(x-x_a),
    \qquad
    \int d^d x\,\nu_{X_N}(x)=1,
    \label{eq:empirical_density}
\end{equation}
and its smeared form
\begin{equation}
    \nu_y[X_N]
    =
    \int d^d x\,\nu_{X_N}(x)L_y(x).
    \label{eq:smeared_empirical_density}
\end{equation}
Then
\begin{equation}
    \Delta N_y^{(N)}(X_N,t) =
    N\bigl[\nu_y[X_N]-\bar\nu_y(t)\bigr],
\end{equation}
and
\begin{equation}
    \bar\nu_y(t) =
    \int dX_N\,\rho_N(X_N,t)\nu_y[X_N].
    \label{eq:Delta_N_empirical}
\end{equation}
The stochastic and deterministic coefficients in
Eq.~\eqref{eq:CSL_fixed_N} therefore contain explicit factors $N$ and
$N^2$, respectively. Their physical effect, however, depends on the
spatial distinguishability of the configurations, as made explicit
below.

\subsection{Configuration-space decoherence and amplification}
\label{subsec:minimal_amplification}

To make the amplification mechanism explicit, let
\begin{equation}
    \overline{\varrho}_N(X_N,X_N',t)
    =
    \mathbb E\!\left[
    \psi_N(X_N,t)\psi_N^*(X_N',t)
    \right]
    \label{eq:rhoN_ensemble}
\end{equation}
be the ensemble-averaged density matrix in the $N$-particle sector.
Equation~\eqref{eq:CSL_fixed_N} implies
\begin{align}
    \partial_t\overline{\varrho}_N(X_N,X_N',t)
    &={}
    -i\bigl[H_N(X_N)-H_N(X_N')\bigr]
    \nonumber\\
    &\times
    \overline{\varrho}_N(X_N,X_N',t)
    \nonumber\\
    &
    -\frac{\lambda^2}{2}
    \mathcal D_N(X_N,X_N')
    \overline{\varrho}_N(X_N,X_N',t),
    \label{eq:CSL_density_matrix_fixed_N}
\end{align}
where the configuration-space decoherence functional is
\begin{equation}
    \mathcal D_N(X_N,X_N')
    =
    \int d^d y\,
    \bigl[N_y^{(N)}(X_N)-N_y^{(N)}(X_N')\bigr]^2.
    \label{eq:decoherence_functional}
\end{equation}
Using Eq.~\eqref{eq:kernel_definition}, this becomes
\begin{align}
    \mathcal D_N(X_N,X_N')
    =
    \sum_{a,b=1}^{N}
    \Bigl[
    &g(x_a-x_b)+g(x_a'-x_b')
    \nonumber\\
    &-2g(x_a-x_b')
    \Bigr].
    \label{eq:decoherence_kernel_sum}
\end{align}
This expression, rather than the formal power of $N$ in the stochastic equation alone, determines the suppression rate of a
spatial superposition. The internal pair structure of a single configuration is
\begin{equation}
    \int d^d y\,
    \bigl[N_y^{(N)}(X_N)\bigr]^2
    =
    Ng(0)+2\sum_{a<b}g(x_a-x_b).
    \label{eq:self_pair_decomposition}
\end{equation}
The first term contains the $N$ diagonal self-contractions, whereas
the second contains the off-diagonal pair contributions. Consider,
for example, two spatially compact configurations related by a rigid
translation, $x_a'=x_a+R$, with all internal separations much smaller
than $r_C$. In this regime,
\begin{equation}
    \mathcal D_N(X_N,X_N')
    \simeq
    2N^2\bigl[g(0)-g(R)\bigr].
    \label{eq:compact_amplification}
\end{equation}
Thus well-separated compact branches exhibit the characteristic
$N^2$ CSL amplification. For dilute configurations, off-diagonal
terms with $|x_a-x_b|\gg r_C$ are suppressed, and the decoherence rate
reduces to a sum over particles or spatially correlated blocks. The
amplification is therefore controlled jointly by particle number and
by the mass distribution relative to the correlation length.

\subsection{Probability balance, phase dynamics, and equivariance}
\label{subsec:minimal_probability}

Write
\begin{equation}
    \psi_N(X_N,t)
    =
    \sqrt{\rho_N(X_N,t)}\,e^{iS_N(X_N,t)}.
    \label{eq:polar_minimal}
\end{equation}
Applying It\^o's rule,
$d\rho_N=\psi_N^*d\psi_N+\psi_Nd\psi_N^*+d\psi_N^*d\psi_N$,
the deterministic localization contribution cancels against the
quadratic It\^o term. One obtains
\begin{equation}
    d\rho_N
    +
    \sum_{a=1}^{N}\nabla_a\cdot J_a\,dt
    =
    2\rho_N\,d\Omega_N,
    \label{eq:CSL_probability_balance}
\end{equation}
where
\begin{align}
    J_a(X_N,t)
    & =
    \frac{\rho_N}{m}\nabla_aS_N,
    \nonumber\\
    d\Omega_N(X_N,t)
    & =
    \lambda\int d^d y\,
    \Delta N_y^{(N)}(X_N,t)dW_t(y).
    \label{eq:minimal_current_noise}
\end{align}
For a normalized fixed-sector state,
\begin{equation}
    \int dX_N\,\rho_N(X_N,t)
    \Delta N_y^{(N)}(X_N,t)=0,
    \label{eq:centered_identity}
\end{equation}
so that Eq.~\eqref{eq:CSL_probability_balance} preserves the total
probability for each noise realization under the usual vanishing-flux
boundary conditions.

Because the minimal CSL operator is real and multiplicative in
configuration space, it does not generate a direct stochastic phase
increment. The Hamilton--Jacobi equation retains the form
\begin{equation}
    dS_N
    =
    -\left[
    \sum_{a=1}^{N}\frac{(\nabla_aS_N)^2}{2m}
    +Q_N
    \right]dt,
    \label{eq:minimal_HJ}
\end{equation}
with
\begin{equation}
    Q_N
    =
    -\sum_{a=1}^{N}
    \frac{1}{2m}
    \frac{\nabla_a^2\sqrt{\rho_N}}{\sqrt{\rho_N}}.
    \label{eq:minimal_quantum_potential}
\end{equation}
The phase is nevertheless affected indirectly because the stochastic
evolution of $\rho_N$ changes $Q_N$. This provides a useful contrast
with the dissipative model, whose current-dependent collapse operator
produces direct stochastic and deterministic phase corrections.

For a fixed realization of the CSL noise, an ensemble of trajectories
transported by the standard Bohmian velocity
\begin{equation}
    v_a(X_N,t)=\frac{\nabla_aS_N}{m}
    \label{eq:standard_Bohm_velocity}
\end{equation}
would have a distribution $f_N$ satisfying
\begin{equation}
    \partial_t f_N
    +
    \sum_{a=1}^{N}\nabla_a\cdot(f_Nv_a)=0.
    \label{eq:trajectory_transport}
\end{equation}
This transport equation does not contain the stochastic source in
Eq.~\eqref{eq:CSL_probability_balance}. Consequently,
$f_N(t_0)=\rho_N(t_0)$ does not in general imply
$f_N(t)=\rho_N(t)$ for the same noise realization. The standard
Bohmian flow is therefore not pathwise equivariant under minimal CSL.
The ordinary continuity equation is recovered exactly for
$\lambda\to0$, and approximately in regimes where the CSL
contribution is negligible.

\subsection{Energy growth}
\label{subsec:minimal_energy}

The localization dynamics also induces momentum diffusion. Let
$E_N(t)=\langle H_N\rangle_t$. It\^o calculus applied to
Eq.~\eqref{eq:CSL_fixed_N} gives
\begin{equation}
    dE_N
    =
    dE_N^{\rm st}
    +
    \frac{\lambda^2}{2}
    \int d^d y\,
    \left\langle
    \left[
    N_y^{(N)},
    \left[H_N,N_y^{(N)}\right]
    \right]
    \right\rangle_tdt,
    \label{eq:EN_balance_commutator}
\end{equation}
where $dE_N^{\rm st}$ is linear in the Wiener increments and satisfies
$\mathbb E[dE_N^{\rm st}]=0$. For the free Hamiltonian,
\begin{equation}
    \left[
    N_y^{(N)},
    \left[H_N,N_y^{(N)}\right]
    \right]
    =
    \frac{1}{m}
    \sum_{a=1}^{N}|\nabla_aL_y(x_a)|^2.
    \label{eq:minimal_double_commutator}
\end{equation}
Translational invariance gives
\begin{equation}
    C_{ii}
    \equiv
    \int d^d y\,|\nabla L_y(x)|^2
    =
    -\nabla^2g(x)\big|_{x=0},
    \label{eq:Cii_definition}
\end{equation}
and hence
\begin{equation}
    \frac{d}{dt}\mathbb E[E_N(t)]
    =
    P_N^{\rm CSL},
    \qquad
    P_N^{\rm CSL}
    =
    N\frac{\gamma m}{2m_0^2}C_{ii}.
    \label{eq:minimal_heating_rate}
\end{equation}
Therefore
\begin{equation}
    \mathbb E[E_N(t)]
    =
    \mathbb E[E_N(0)]+P_N^{\rm CSL}t.
    \label{eq:CSL_heating}
\end{equation}
For the Gaussian kernel
\begin{equation}
    g(r)=g(0)e^{-r^2/(4r_C^2)},
    \label{eq:Gaussian_kernel_minimal}
\end{equation}
one has $C_{ii}=dg(0)/(2r_C^2)$. Restoring $\hbar$, the
ensemble-averaged heating rate becomes
\begin{equation}
    \frac{d}{dt}\mathbb E[E_N(t)]
    =
    N\frac{d\,\gamma g(0)\hbar^2m}{4m_0^2r_C^2},
    \label{eq:standard_CSL_heating}
\end{equation}
which is the standard CSL result. The linear $N$ dependence of the
kinetic-energy growth should be distinguished from the $N^2$
decoherence amplification of compact configurations: the double
commutator retains only same-particle contractions. Since the minimal
model contains no friction term proportional to $-E_N$, it has no
finite asymptotic equilibrium energy.

\subsection{Fock-sector weights and coherences}
\label{subsec:minimal_sector_superpositions}

For a general Fock-space state, Eq.~\eqref{eq:CSL_functional} remains
diagonal in particle number, but its nonlinear centering term contains
the total expectation value
\begin{equation}
    \bar N_y^{\rm tot}(t)
    =
    \sum_{N=0}^{\infty}
    \int dX_N\,\rho_N(X_N,t)N_y^{(N)}(X_N).
    \label{eq:total_N_mean}
\end{equation}
Thus an initially absent sector is not generated, although the
stochastic evolution of every occupied sector depends on the same
global centering term. The fixed-sector equation
\eqref{eq:CSL_fixed_N} applies with
$\bar N_y(t)$ replaced by $\bar N_y^{\rm tot}(t)$.

Define the sector weight
\begin{equation}
    P_N(t)
    =
    \int dX_N\,\rho_N(X_N,t),
    \qquad
    \sum_{N=0}^{\infty}P_N(t)=1,
    \label{eq:sector_weights}
\end{equation}
and, for $P_N(t)\neq0$, the conditional sector expectation
\begin{equation}
    \bar N_y^{(N)}(t)
    =
    \frac{1}{P_N(t)}
    \int dX_N\,\rho_N(X_N,t)N_y^{(N)}(X_N).
    \label{eq:conditional_sector_mean}
\end{equation}
Integration over the $N$-particle configuration space gives
\begin{equation}
    dP_N(t)
    =
    2\lambda P_N(t)
    \int d^d y\,
    \bigl[\bar N_y^{(N)}(t)-\bar N_y^{\rm tot}(t)\bigr]
    dW_t(y).
    \label{eq:PN_balance}
\end{equation}
The relative weights therefore fluctuate along individual noise
realizations, while
\begin{equation}
    \mathbb E[P_N(t)]=P_N(0)
    \label{eq:PN_martingale}
\end{equation}
shows that each sector weight is a martingale. Summing
Eq.~\eqref{eq:PN_balance} over $N$ gives pathwise conservation of the
total norm.

The suppression of coherences is most directly expressed by the
ensemble-averaged off-diagonal block
\begin{equation}
    \overline{\varrho}_{NM}(X_N,X_M',t)
    =
    \mathbb E\!\left[
    \psi_N(X_N,t)\psi_M^*(X_M',t)
    \right].
    \label{eq:rho_NM_definition}
\end{equation}
It satisfies
\begin{align}
    \partial_t\overline{\varrho}_{NM}
    ={}&
    -i\bigl[H_N(X_N)-H_M(X_M')\bigr]
    \overline{\varrho}_{NM}
    \nonumber\\
    &
    -\frac{\lambda^2}{2}
    \mathcal D_{NM}(X_N,X_M')
    \overline{\varrho}_{NM},
    \label{eq:rho_NM_evolution}
\end{align}
where
\begin{equation}
    \mathcal D_{NM}(X_N,X_M')
    =
    \int d^d y\,
    \bigl[N_y^{(N)}(X_N)-N_y^{(M)}(X_M')\bigr]^2.
    \label{eq:decoherence_NM}
\end{equation}
Hence CSL suppresses coherences only to the extent that the two
sectors carry distinguishable smeared mass-density profiles. It does
not directly measure the abstract particle-number label and does not
create or annihilate particles. Under persistent distinguishability,
the stochastic competition among the weights may lead to effective
sector selection. In a strictly isolated nonrelativistic theory, the
operational meaning of coherences between sectors with different
total masses requires additional care; here they are treated formally
as Fock-space coherences.

\section{Dissipative CSL in fixed particle-number sectors}
\label{sec:dissipative_csl}

We now consider the dissipative extension in which the smeared mass-density
collapse operator is supplemented by a current-dependent contribution. The
one-particle sector provides a consistency check against the known balance
between collapse-induced heating and friction. The fixed-$N$ projection then
reveals structures that have no one-particle analogue: a deterministic pair
drift in the probability balance, correlated diffusion in configuration space,
mixed phase-curvature terms, and an off-diagonal contribution to the kinetic-energy
balance. We introduce only the additional parameters
\begin{equation}
\kappa=\frac{\hbar^2\beta}{4},
\qquad
\chi=\frac{\kappa}{m r_C^2},
\label{eq:chi_definition_consistent}
\end{equation}
while the CSL parameters and the kernel conventions are those of
Sec.~\ref{sec:minimal_CSL}.

\subsection{Functional collapse operator and fixed-sector projection}
\label{subsec:functional_dissipative_operator}

Momentum-dependent dissipative CSL models were introduced in
Refs.~\cite{Smirne2014,Smirne2015}. Here we adopt the linear-friction
many-body construction of Ref.~\cite{DiBartolomeo2023}, in which the Hermitian
mass-density operator is supplemented by the divergence of the current operator.
Recent phenomenological bounds on this construction were discussed in
Ref.~\cite{DiBartolomeo2024}. In the Bargmann functional representation, the
collapse operator is
\begin{equation}
\mathcal L_y
=
\mathcal M_y
-i\frac{\hbar\beta}{4}\,
\nabla_y\cdot\boldsymbol{\mathcal J}_y,
\label{eq:dissipative_operator_functional_revised}
\end{equation}
where $\mathcal M_y=m\mathcal N_y$ and
\begin{equation}
\begin{aligned}
\boldsymbol{\mathcal J}_y
={}&
-\frac{i\hbar}{2}
\int d^d x\,L_y(x)
\Bigg[
\bar\alpha(x)\nabla_x
\frac{\delta}{\delta\bar\alpha(x)}
\\
&\hspace{26mm}
-
\bigl(\nabla_x\bar\alpha(x)\bigr)
\frac{\delta}{\delta\bar\alpha(x)}
\Bigg].
\end{aligned}
\label{eq:functional_current_revised}
\end{equation}
For $\beta=0$, $\mathcal L_y$ reduces to the Hermitian mass-density operator.
For $\beta\neq0$, it is non-Hermitian, and the normalized It\^o equation reads
\begin{equation}
\begin{aligned}
d\Psi[\bar\alpha,t]
={}&
\Bigg[
-i\mathcal H_0\,dt
+
\frac{\sqrt\gamma}{m_0}
\int d^d y\,
\bigl(\mathcal L_y-r_t(y)\bigr)dW_t(y)
\\
&-
\frac{\gamma}{2m_0^2}
\int d^d y\,
\left(
\mathcal L_y^\dagger\mathcal L_y
+r_t^2(y)
-2r_t(y)\mathcal L_y
\right)dt
\Bigg]
\\[-1mm]
&\hspace{8mm}\times\Psi[\bar\alpha,t].
\end{aligned}
\label{eq:dissipative_functional_equation_revised}
\end{equation}
The real centering term is
\begin{equation}
r_t(y)
=
\frac12\left\langle
\mathcal L_y^\dagger+\mathcal L_y
\right\rangle_t
=
\langle\mathcal M_y\rangle_t,
\label{eq:rt_mass_revised}
\end{equation}
because the current contribution is anti-Hermitian under the boundary
conditions assumed throughout.

Projection onto a normalized fixed-$N$ sector gives
\begin{equation}
\begin{aligned}
d\psi_N
={}&
\Bigg[
-iH_Ndt
+
\frac{\sqrt\gamma}{m_0}
\int d^d y\,
\bigl(\ell_y^{(N)}-r_y\bigr)dW_t(y)
\\
&-
\frac{\gamma}{2m_0^2}
\int d^d y\,
\left(
\ell_y^{(N)\dagger}\ell_y^{(N)}
+r_y^2
-2r_y\ell_y^{(N)}
\right)dt
\Bigg]
\\[-1mm]
&\hspace{8mm}\times\psi_N,
\end{aligned}
\label{eq:dissipative_fixed_N_revised}
\end{equation}
with
\begin{align}
\ell_y^{(N)}
&=
\sum_{a=1}^N\left[mL_y(x_a)+\kappa A_{a,y}\right],
\nonumber\\
A_{a,y}
&=
\nabla_aL_y(x_a)\cdot\nabla_a
+
\frac12\nabla_a^2L_y(x_a).
\label{eq:ell_N_revised}
\end{align}
The first-order operator $A_{a,y}$ is anti-Hermitian. Using the empirical
density introduced in Sec.~\ref{sec:minimal_CSL}, one obtains
\begin{equation}
\ell_y^{(N)}-r_y
=
mN\bigl[\nu_y[X_N]-\bar\nu_y(t)\bigr]
+
\kappa\sum_{a=1}^N A_{a,y}.
\label{eq:ell_minus_r_revised}
\end{equation}
Thus the density part retains the usual CSL amplification, whereas the current
part acts through gradients on the sector wave function. Since all operators in
Eq.~\eqref{eq:dissipative_operator_functional_revised} are number-conserving
bilinears, the dissipative dynamics does not create or annihilate particles.
For superpositions of different particle-number sectors, the centering remains
controlled by the total smeared mass density, as in Sec.~\ref{sec:minimal_CSL};
the current term changes the sector profiles but not the sectorial closure.

\subsection{One-particle sector as a consistency check}
\label{subsec:one_particle_consistency}

For one particle,
\begin{equation}
\ell_y^{(1)}
=
mL_y(x)+\kappa A_y,
\qquad
A_y
=
\nabla L_y(x)\cdot\nabla
+
\frac12\nabla^2L_y(x).
\label{eq:ell_one_particle_revised}
\end{equation}
We introduce the smeared noise and the kernel-gradient contraction
\begin{align}
d\mathcal W_t(x)
&=
\int d^d y\,L_y(x)dW_t(y),
\nonumber\\
C_{ij}(x,z)
&=
\int d^d y\,
\partial_iL_y(x)\partial_jL_y(z)
\nonumber\\
&=
-\partial_i\partial_j g(x-z).
\label{eq:Cij_kernel_revised}
\end{align}
so that
$\mathbb E[d\mathcal W_t(x)d\mathcal W_t(z)]=g(x-z)dt$.
For coincident arguments we write $C_{ij}\equiv C_{ij}(x,x)$, which is
independent of $x$ by translational invariance.

Writing $\psi_1=\sqrt{\rho_1}\,e^{iS_1}$, the density equation is
\begin{equation}
\begin{aligned}
d\rho_1
={}&
-\nabla_i\left[
\frac{\rho_1}{m}
\left(
\nabla_iS_1
-
\frac{\lambda^2\kappa^2}{2m}
C_{ij}\nabla_j\ln\rho_1
\right)
\right]dt
\\
&+
2\rho_1d\Omega_1
+
\frac{\lambda\kappa}{m}
\nabla_i\left[
\rho_1\nabla_i d\mathcal W_t(x)
\right],
\end{aligned}
\label{eq:drho1_revised}
\end{equation}
where $d\Omega_1$ is the minimal-CSL increment defined in
Sec.~\ref{sec:minimal_CSL}. The corresponding phase equation is more
transparently written without introducing an effective quantum potential:
\begin{equation}
\begin{aligned}
dS_1
={}&
\left[
-\frac{(\nabla S_1)^2}{2m}
-Q_1
\right]dt
\\
&+
\frac{2\lambda^2\kappa}{m}
\nabla_iS_1\nabla_i\mathcal G_1\,dt
\\
&+
\frac{\lambda^2\kappa^2}{2m^2}
C_{ij}\nabla_i\nabla_jS_1\,dt
\\
&+
\frac{\lambda\kappa}{m}
\nabla_iS_1\nabla_i d\mathcal W_t(x).
\end{aligned}
\label{eq:dS1_revised}
\end{equation}
with
\begin{align}
Q_1
&=
-\frac{1}{2m}
\frac{\nabla^2\sqrt{\rho_1}}{\sqrt{\rho_1}},
\nonumber\\
\mathcal G_1(x,t)
&=
\int d^d z\,\rho_1(z,t)g(z-x).
\label{eq:G1_Q1_revised}
\end{align}
Equations~\eqref{eq:drho1_revised} and \eqref{eq:dS1_revised} show the
contrast with minimal CSL: the current-dependent collapse operator generates
both a stochastic probability current and a direct stochastic phase increment.
The explicit derivation will be given in
Appendix~\ref{app:one_particle_dissipative_projection}.

The ensemble-averaged kinetic energy obeys
\begin{equation}
\frac{d}{dt}\mathbb E[\mathcal E_1]
=
P_{\rm eff}^{(1)}
-
\Gamma_{\rm eff}^{(1)}\mathbb E[\mathcal E_1],
\label{eq:one_particle_energy_balance_revised}
\end{equation}
and the Gaussian-kernel coefficients are collected in
Appendix~\ref{app:energy_balance_commutators}. When
$\Gamma_{\rm eff}^{(1)}>0$, the stationary energy may be parametrized by
\begin{equation}
k_BT_{\rm eff}^{(d)}
=
\frac{1}{4\kappa}
\frac{
1-\dfrac{d+2}{2}\chi
+\dfrac{(d+2)(d+4)}{16}\chi^2
}{
1-\dfrac{3(d+2)}{8}\chi
}.
\label{eq:Teff_revised}
\end{equation}
For $\kappa>0$, positive damping requires
\begin{equation}
0<\chi<\frac{8}{3(d+2)}.
\label{eq:chi_damping_domain}
\end{equation}
This one-particle result is a consistency check rather than a new prediction.

\subsection{Many-body probability and phase dynamics}
\label{subsec:fixed_N_collective_pair_terms}

Let $\psi_N=\sqrt{\rho_N}\,e^{iS_N}$ and define
$p_{a,i}=\nabla_{a,i}S_N$. Expanding Eq.~\eqref{eq:dissipative_fixed_N_revised}
and collecting the terms in divergence form gives
\begin{equation}
\begin{aligned}
d\rho_N
&+
\sum_{a=1}^N
\nabla_a\cdot\left(
\frac{\rho_N\nabla_aS_N}{m}
\right)dt
=
2\rho_Nd\Omega_N
\\
&+
\frac{\lambda\kappa}{m}
\sum_{a=1}^N
\nabla_a\cdot\left[
\rho_N\nabla_a d\mathcal W_t(x_a)
\right]
\\
&+
\frac{\lambda^2\kappa}{m}
\sum_{a\neq b}
\nabla_a\cdot\left[
\rho_N\nabla_a g(x_a-x_b)
\right]dt
\\
&+
\frac{\lambda^2\kappa^2}{2m^2}
\sum_{a,b=1}^N
\partial_{a i}\partial_{b j}
\left[
\rho_N C_{ij}(x_a,x_b)
\right]dt.
\end{aligned}
\label{eq:drhoN_dissipative_revised}
\end{equation}
Here $d\Omega_N$ is the minimal-CSL multiplicative noise. The third line is a
deterministic pair drift of order $\kappa$, whereas the last line is a
configuration-space diffusion term whose off-diagonal components correlate
different particle coordinates. Both contributions are suppressed when the
relevant separations are much larger than $r_C$. Equation
\eqref{eq:drhoN_dissipative_revised} also makes norm preservation explicit:
all current-dependent terms integrate to boundary terms, while the centered
minimal-CSL contribution integrates to zero.

The deterministic phase-gradient term is controlled by
\begin{equation}
\mathcal G_N(x;X_N,t)
=
\int d^d z\,
\bigl[2\bar\nu(z,t)-\nu_{X_N}(z)\bigr]g(z-x),
\label{eq:GN_revised}
\end{equation}
and it is convenient to define the shifted phase gradient
\begin{align}
\Pi_{a i}^{(N)}
&=
p_{a,i}
-
\eta_N\nabla_i\mathcal G_N(x_a;X_N,t),
\nonumber\\
\eta_N
&=
\frac{\gamma\kappa m^2N}{m_0^2}
=
\lambda^2\kappa N.
\label{eq:PiN_revised}
\end{align}
The full phase equation can be written as
\begin{equation}
\begin{aligned}
dS_N
={}&
\Bigg[
-\sum_{a=1}^N\frac{p_a^2}{2m}
-Q_N
+
\frac{\eta_N}{m}
\sum_{a=1}^N
p_a\cdot\nabla\mathcal G_N(x_a)
\\
&\hspace{8mm}
+
\frac{\gamma\kappa^2}{2m_0^2}
\sum_{a,b=1}^N
C_{ij}(x_a,x_b)
\partial_{a i}\partial_{b j}S_N
\Bigg]dt
\\
&+
\frac{\lambda\kappa}{m}
\sum_{a=1}^N
p_a\cdot\nabla_a d\mathcal W_t(x_a),
\end{aligned}
\label{eq:dSN_full_revised}
\end{equation}
where
\begin{equation}
Q_N
=
-\sum_{a=1}^N
\frac{1}{2m}
\frac{\nabla_a^2\sqrt{\rho_N}}{\sqrt{\rho_N}}.
\end{equation}
Equivalently, the deterministic part may be reorganized as
\begin{equation}
\begin{aligned}
(dS_N)_{\rm det}
={}&
\Bigg[
-\sum_{a=1}^N
\frac{\bigl(\Pi_a^{(N)}\bigr)^2}{2m}
-Q_N
+
\frac{\eta_N^2}{2m}
\sum_{a=1}^N
\left|\nabla\mathcal G_N(x_a)\right|^2
\\
&\hspace{4mm}
+
\frac{\gamma\kappa^2}{2m_0^2}
\sum_{a,b=1}^N
C_{ij}(x_a,x_b)
\partial_{a i}\partial_{b j}S_N
\Bigg]dt.
\end{aligned}
\label{eq:dSN_shifted_revised}
\end{equation}
The off-diagonal phase-curvature contribution is
\begin{equation}
\mathcal R_{\rm mix}^{(N)}
=
\frac{\gamma\kappa^2}{2m_0^2}
\sum_{a\neq b}
C_{ij}(x_a,x_b)
\partial_{a i}\partial_{b j}S_N.
\label{eq:Rmix_revised}
\end{equation}
It vanishes for $N=1$ and is exponentially suppressed in the dilute regime.
Compactness prevents kernel suppression but does not, by itself, guarantee a
nonzero contribution: $\mathcal R_{\rm mix}^{(N)}$ also requires mixed phase
curvatures. In particular, it vanishes for an additive phase
$S_N(X_N)=\sum_aS_a(x_a)$.

The notation $\Pi_a^{(N)}$ in Eq.~\eqref{eq:PiN_revised} denotes an algebraically
shifted phase gradient. It should not be identified with a probability-transport
momentum. Equation~\eqref{eq:drhoN_dissipative_revised} contains pair drifts,
second-order diffusion, and stochastic source terms that are not generated by the
flow $\Pi_a^{(N)}/m$. Consequently, the standard Bohmian flow is not exactly
equivariant for a fixed noise realization, and the Hamilton--Jacobi completion of
squares does not single out a canonical modified guidance equation.

\subsection{Operator energy balance and compact/dilute limits}
\label{subsec:energy_balance_revised}

Let $\mathcal E_N(t)=\langle H_N\rangle_t$. It\^o calculus gives
\begin{equation}
\begin{aligned}
d\mathcal E_N
={}&
d\mathcal E_N^{\rm st}
+
\frac{\gamma}{2m_0^2}
\int d^d y\,
\Big\langle
\ell_y^{(N)\dagger}[H_N,\ell_y^{(N)}]
\\
&\hspace{25mm}
+[\ell_y^{(N)\dagger},H_N]\ell_y^{(N)}
\Big\rangle_tdt.
\end{aligned}
\label{eq:dEN_general_revised}
\end{equation}
The stochastic term has zero ensemble mean. Separating diagonal and off-diagonal
contractions yields
\begin{equation}
\boxed{\frac{d}{dt}\mathbb E[\mathcal E_N]
=
NP_{\rm eff}^{(1)}
-
\Gamma_{\rm eff}^{(1)}\mathbb E[\mathcal E_N]
+
\mathbb E[\mathcal C_N],}
\label{eq:EN_balance_revised}
\end{equation}
where the pair contribution must first be kept in operator form,
\begin{equation}
\mathcal C_N
=
\frac{\gamma\kappa}{m_0^2}
\sum_{a<b}
\left\langle
 g_{ij}(r_{ab})
 \widehat P_{ab,i}\widehat P_{ab,j}
\right\rangle_t,
\label{eq:CN_revised}
\end{equation}
with
\begin{align}
r_{ab}
&=x_a-x_b,
\qquad
g_{ij}(r)=\partial_i\partial_jg(r),
\nonumber\\
\widehat{\boldsymbol P}_{ab}
&=
\widehat{\boldsymbol p}_a+\widehat{\boldsymbol p}_b
=
-i(\nabla_a+\nabla_b).
\label{eq:Pab_operator_revised}
\end{align}
Because $(\nabla_a+\nabla_b)g(r_{ab})=0$, the kernel commutes with the
total momentum of the pair. Therefore Eq.~\eqref{eq:CN_revised} is Hermitian and
its polar representation is
\begin{equation}
\begin{aligned}
\mathcal C_N
={}&
\frac{\gamma\kappa}{m_0^2}
\sum_{a<b}
\int dX_N\,\rho_N g_{ij}(r_{ab})
\Big[
D_{ab,i}S_ND_{ab,j}S_N
\\
&\hspace{1.cm}
+D_{ab,i}\ln\sqrt{\rho_N}\,
 D_{ab,j}\ln\sqrt{\rho_N}
\Big],\\
\end{aligned}
\label{eq:CN_polar_revised}
\end{equation}
where $D_{ab}=\nabla_a+\nabla_b$. The second term in
Eq.~\eqref{eq:CN_polar_revised} is an amplitude contribution and cannot in general
be discarded.

For a Gaussian kernel,
\begin{equation}
g_{ij}(r)
=
\left(
\frac{r_ir_j}{4r_C^4}
-
\frac{\delta_{ij}}{2r_C^2}
\right)g(r).
\label{eq:gij_gaussian_revised}
\end{equation}
The tensor $g_{ij}(r)$ is not sign-definite at arbitrary separation, so
$\mathcal C_N$ cannot generally be interpreted as friction. In the dilute regime,
$r_{ab}\gg r_C$, it is exponentially suppressed and
\begin{equation}
\mathcal C_N\simeq0,
\qquad
\mathcal E_{N,{\rm eq}}\simeq N\mathcal E_{1,{\rm eq}}.
\label{eq:dilute_energy_balance_revised}
\end{equation}
In the compact regime, $r_{ab}\ll r_C$,
\begin{equation}
g_{ij}(r_{ab})
\simeq
-\frac{g(0)}{2r_C^2}\delta_{ij},
\end{equation}
and hence
\begin{equation}
\mathcal C_N
\simeq
-\mu_C
\sum_{a<b}
\left\langle
\widehat{\boldsymbol P}_{ab}^{\,2}
\right\rangle_t,
\qquad
\mu_C
=
\frac{\gamma\kappa g(0)}{2m_0^2r_C^2}.
\label{eq:CN_compact_revised}
\end{equation}
Using the operator identity
\begin{equation}
\sum_{a<b}
\widehat{\boldsymbol P}_{ab}^{\,2}
=
(N-2)\sum_{a=1}^N\widehat{\boldsymbol p}_a^{\,2}
+
\widehat{\boldsymbol P}_{\rm tot}^{\,2},
\qquad
\widehat{\boldsymbol P}_{\rm tot}
=
\sum_{a=1}^N\widehat{\boldsymbol p}_a,
\label{eq:pair_identity_revised}
\end{equation}
one obtains
\begin{equation}
\begin{aligned}
\frac{d}{dt}\mathbb E[\mathcal E_N]
={}&
NP_{\rm eff}^{(1)}
-
\left[
\Gamma_{\rm eff}^{(1)}
+2m\mu_C(N-2)
\right]
\mathbb E[\mathcal E_N]
\\
&-
\mu_C\,
\mathbb E\!\left[
\left\langle
\widehat{\boldsymbol P}_{\rm tot}^{\,2}
\right\rangle_t
\right]
\end{aligned}
\label{eq:compact_balance_with_Ptot_revised}
\end{equation}
to leading compact order. Notice that
$\langle\widehat{\boldsymbol P}_{\rm tot}\rangle=0$ is not sufficient to remove
the last term; its dispersion must also be negligible.

When
\begin{equation}
\mathbb E\!\left[
\left\langle
\widehat{\boldsymbol P}_{\rm tot}^{\,2}
\right\rangle_t
\right]
\ll
2m(N-2)\mathbb E[\mathcal E_N]
\label{eq:Ptot_closure_condition_revised}
\end{equation}
for $N>2$, the balance closes as
\begin{equation}
\frac{d}{dt}\mathbb E[\mathcal E_N]
=
NP_{\rm eff}^{(1)}
-
\Gamma_N^{\rm comp}\mathbb E[\mathcal E_N],
\label{eq:compact_balance_revised}
\end{equation}
where, for the Gaussian kernel,
\begin{align}
\Gamma_N^{\rm comp}
&=
\frac{\gamma\kappa m C_d}{m_0^2}
\left[
2N-\frac{3(d+2)}{2}\chi
\right],
\nonumber\\
C_d
&=
\frac{1}{2^{d+1}\pi^{d/2}r_C^{d+2}}.
\label{eq:Gamma_compact_revised}
\end{align}
The corresponding energy and its temperature parametrization are
\begin{equation}
\mathcal E_{N,{\rm eq}}^{\rm comp}
=
\frac{NP_{\rm eff}^{(1)}}{\Gamma_N^{\rm comp}}
=
N\mathcal E_{1,{\rm eq}}
\frac{\Gamma_{\rm eff}^{(1)}}{\Gamma_N^{\rm comp}},
\label{eq:EN_compact_eq_revised}
\end{equation}
and
\begin{equation}
T_N^{\rm comp}
=
T_{\rm eff}
\frac{
1-\dfrac{3(d+2)}{8}\chi
}{
\dfrac{N}{2}-\dfrac{3(d+2)}{8}\chi
}.
\label{eq:TN_compact_revised}
\end{equation}
This temperature only parametrizes the leading compact energy balance; it does not
imply relaxation to a Gibbs state. Beyond the compact approximation, the pair term
depends on spatial and momentum correlations and the balance does not close in terms
of $\mathbb E[\mathcal E_N]$ alone. For $N=2$, the internal-energy contribution in
Eq.~\eqref{eq:pair_identity_revised} vanishes at leading compact order, but the term
$\langle\widehat{\boldsymbol P}_{\rm tot}^{\,2}\rangle_t$ remains unless the state
is restricted to a sharply defined zero-total-momentum sector.


\subsection{Correlated Gaussian internal states beyond the compact limit}
\label{sec:gaussian_internal}

To illustrate the many-body terms without imposing the compact
approximation, we consider a translationally invariant Gaussian state in
internal coordinates. Defining
\begin{equation}
 \mathbf R=\frac{1}{N}\sum_{a=1}^{N}\mathbf x_a,\qquad
 \mathbf y_a=\mathbf x_a-\mathbf R,
\end{equation}
and
\begin{equation}
 Q=\sum_{a=1}^{N}\mathbf y_a^2,\qquad
 \sum_{a=1}^{N}\mathbf y_a=0,
\end{equation}
we restrict the state to a sharply defined zero-total-momentum sector and use
\begin{equation}
 \psi_N(X_N,t)
 =
 \frac{
 \exp\left[
 -Q/(4\sigma^2)
 +ibQ/2+i\theta
 \right]
 }{
 [2\pi\sigma^2]^{d(N-1)/4}
 }.
 \label{eq:internal_gaussian}
\end{equation}
Here, $\sigma(t)$ is the internal width and $b(t)$ determines the
quadratic phase. The corresponding kinetic energy is
\begin{equation}
 \mathcal E_N(\sigma,b)
 =
 d(N-1)
 \left[
 \frac{1}{8m\sigma^2}
 +\frac{b^2\sigma^2}{2m}
 \right].
 \label{eq:gaussian_energy}
\end{equation}

For the full Gaussian CSL kernel,
\begin{equation}
 g(r)=g(0)e^{-r^2/(4r_C^2)},
\end{equation}
all pair averages can be evaluated without expanding in $r/r_C$. In
particular, introducing
\begin{equation}
 u=\frac{\sigma^2}{r_C^2},
 \qquad
 {\cal F}_d(u)=(1+u)^{-(d+2)/2},
 \label{eq:gaussian_crossover}
\end{equation}
the collective contribution to the energy balance becomes
\begin{equation}
 \boxed{
 \mathcal C_N
 =
 -2m\mu_C(N-2)\mathcal E_N(\sigma,b)\,
 {\cal F}_d(u)
 },\qquad
 \label{eq:CN_gaussian_exact}
\end{equation}
where $\mu_C=\frac{\gamma\kappa g(0)}
 {2m_0^2r_C^2}$. Thus ${\cal F}_d(u)$ provides an explicit crossover between the compact
and dilute regimes. Equation~\eqref{eq:CN_gaussian_exact} reproduces
$\mathcal C_N=-2m\mu_C(N-2)\mathcal E_N$ for $\sigma\ll r_C$, whereas
$\mathcal C_N\rightarrow0$ for $\sigma\gg r_C$. Moreover,
\begin{equation}
 \mathcal C_2=0
 \label{eq:C2_gaussian_zero}
\end{equation}
for every value of $\sigma/r_C$, and not only in the compact limit. This
exact cancellation follows from
$\widehat{\boldsymbol P}_{12}=\widehat{\boldsymbol P}_{\mathrm{tot}}=0$ in the selected sector. A moment projection of the deterministic probability equation,
using $\langle Q\rangle=d(N-1)\sigma^2$, yields
\begin{equation}
 \begin{split}
 \dot{\sigma}
 ={}&
 \frac{b\sigma}{m}
 +\frac{\eta_N g(0)}{2mr_C^2}
 \sigma\,{\cal F}_d(u)
 \\
 &+
 \frac{\gamma\kappa^2g(0)}
 {4m_0^2r_C^2}
 \sigma\left[1-{\cal F}_d(u)\right],
 \end{split}
 \label{eq:sigma_gaussian_projected}
\end{equation}
where $\eta_N=\gamma\kappa m^2N/m_0^2$. The three terms respectively
describe the Hamiltonian flow, the pair drift linear in $\kappa$, and the
correlated diffusion of order $\kappa^2$.

Equation~\eqref{eq:sigma_gaussian_projected} should not be interpreted
as an exact Gaussian solution of the full fixed-$N$ equation. Away from
the compact regime, the deterministic drift generates structures of the
form
\begin{equation}
 \sum_{a\neq b}
 e^{-r_{ab}^2/(4r_C^2)}
 {\cal P}(\mathbf y_a,\mathbf y_b,\mathbf r_{ab}),
\end{equation}
with ${\cal P}$ a polynomial determined by the kernel derivatives.
These terms cannot, in general, be expressed only through $Q$ and hence
produce non-Gaussian components orthogonal to variations of $\sigma$,
$b$, and $\theta$. The Gaussian equations are therefore projected
evolution equations rather than a pointwise closure.

A further subtlety concerns the averaged density entering $\mathcal G_N$. A
state with exactly vanishing total momentum has a delocalized center of
mass and, in a finite volume $V$, satisfies
\begin{equation}
 \bar{\nu}(\mathbf x)=\frac{1}{V},
 \qquad
 \mathcal G_N(\mathbf x)
 =
 \frac{2}{V}
 -\frac{1}{N}\sum_{c=1}^{N}g(\mathbf x_c-\mathbf x).
 \label{eq:GN_zero_momentum}
\end{equation}
If instead the internal state is conditioned on a fixed center of mass,
the one-particle density is Gaussian with variance
$q\sigma^2$, where $q=(N-1)/N$, and
\begin{equation}
 \mathcal G_N(\mathbf x)
 =
 2h_\sigma(\mathbf x-\mathbf R)
 -\frac{1}{N}\sum_{c=1}^{N}g(\mathbf x_c-\mathbf x),
 \label{eq:GN_pinned}
\end{equation}
with
\begin{equation}
 h_\sigma(\mathbf r)
 =
 \frac{
 e^{-\mathbf r^2/[2(q\sigma^2+2r_C^2)]}
 }{
 [2\pi(q\sigma^2+2r_C^2)]^{d/2}
 }.
\end{equation}
The distinction between Eqs.~\eqref{eq:GN_zero_momentum} and
\eqref{eq:GN_pinned} must be retained when projecting the phase equation.
A systematic treatment of the non-Gaussian residuals, the stochastic
projection, and an independent center-of-mass width is left for future
work.

\subsection{Phenomenological scales and experimental parameter ranges}
\label{sec:phenomenological-scales}

\paragraph{One-particle physical scales.}

The dimensionless parameter $\chi$ introduced in the one-particle
energy balance can be related directly to the temperature parameter
commonly employed in linear-friction collapse models. Restoring
$\hbar$, the current contribution to the collapse operator is
parametrized by
\begin{equation}
\kappa=\frac{\hbar^{2}\beta}{4},
\qquad
\beta=\frac{1}{k_{B}T_{\beta}},
\end{equation}
and therefore
\begin{equation}
\boxed{
\chi
=
\frac{\kappa}{m r_{C}^{2}}
=
\frac{\hbar^{2}}
{4m r_{C}^{2}k_{B}T_{\beta}}
}.
\label{eq:chi-physical}
\end{equation}
Here $T_{\beta}$ is conventionally associated with the temperature
of the collapse field in the linear-friction model
\cite{DiBartolomeo2023,DiBartolomeo2024}. As discussed below, it
should be distinguished from the kinetic-energy scale inferred from
the stationary second moment.

For $d=3$, Eq.~\eqref{eq:appC_Teff} becomes
\begin{equation}
\boxed{
T_{\mathrm{eff}}
=
T_{\beta}
\frac{
1-\frac{5}{2}\chi+\frac{35}{16}\chi^{2}
}{
1-\frac{15}{8}\chi
}
}.
\label{eq:Teff-physical}
\end{equation}
The collapse strength cancels from the ratio
$P_{\mathrm{eff}}^{(1)}/\Gamma_{\mathrm{eff}}^{(1)}$.
Consequently, $\gamma$ controls the relaxation rate but not the
stationary value of the mean kinetic energy.

It is useful to introduce the microscopic temperature scale
\begin{equation}
T_{C}(m,r_{C})
=
\frac{\hbar^{2}}
{4m r_{C}^{2}k_{B}},
\qquad
\chi=\frac{T_{C}}{T_{\beta}}.
\label{eq:temperature-scale}
\end{equation}
Numerically,
\begin{equation}
T_{C}
\simeq
1.21\times10^{-5}\,\mathrm{K}
\left(\frac{m_{0}}{m}\right)
\left(\frac{10^{-7}\,\mathrm{m}}{r_{C}}\right)^{2}.
\label{eq:temperature-scale-numerical}
\end{equation}
For a nucleon with $r_{C}=10^{-7}\,\mathrm{m}$ and
$T_{\beta}\sim0.1$--$10\,\mathrm{K}$, as commonly considered for a
collapse field of possible cosmological origin \cite{Toros2017}, one
has $\chi\sim10^{-4}$--$10^{-6}$. The weak-dissipation expansion is
then
\begin{equation}
T_{\mathrm{eff}}
=
T_{\beta}
\left[
1-\frac{5}{8}\chi+O(\chi^{2})
\right],
\label{eq:Teff-weak}
\end{equation}
so that $T_{\mathrm{eff}}\simeq T_{\beta}$ throughout this range.
Significant finite-$r_{C}$ corrections arise only when $T_{\beta}$
approaches the microscopic scale $T_{C}$.

For $d=3$, positive one-particle damping requires
\begin{equation}
0<\chi<\frac{8}{15},
\qquad
T_{\beta}>
\frac{15\hbar^{2}}
{32m r_{C}^{2}k_{B}}.
\label{eq:chi-stability-3d}
\end{equation}
At the upper boundary, $\Gamma_{\mathrm{eff}}^{(1)}$ vanishes and the
stationary second-moment scale in Eq.~\eqref{eq:Teff-physical}
diverges.

\paragraph{Compact many-body scale.}

The one-particle result does not exhaust the phenomenological
implications of the energy balance. Under the compact closure
condition stated in Eq.~\eqref{eq:Ptot_closure_condition_revised} and the closed balance in
Eq.~\eqref{eq:compact_balance_revised}, the many-body parametrization in
Eq.~\eqref{eq:TN_compact_revised} becomes, for $d=3$,
\begin{equation}
\boxed{
T_{N}^{\mathrm{comp}}
=
T_{\beta}
\frac{
1-\frac{5}{2}\chi+\frac{35}{16}\chi^{2}
}{
\frac{N}{2}-\frac{15}{8}\chi
},
\qquad N>2.
}
\label{eq:Tcomp-physical}
\end{equation}
The apparent one-particle singularity at $\chi=8/15$ cancels in this
expression. Within the leading compact mean-energy closure,
$\Gamma_{N}^{\mathrm{comp}}$ remains positive for
\begin{equation}
0<\chi<\frac{4N}{15}.
\label{eq:compact-stability}
\end{equation}
Thus the collective pair contribution extends the domain of positive
mean-energy damping beyond the corresponding one-particle condition.
This statement concerns the closed equation for the mean kinetic
energy only and does not guarantee the closure or finiteness of
higher energy moments.

For weak dissipation,
\begin{equation}
T_{N}^{\mathrm{comp}}
=
\frac{2T_{\beta}}{N}
\left[
1-
\left(
\frac{5}{2}-\frac{15}{4N}
\right)\chi
+O(\chi^{2})
\right].
\label{eq:Tcomp-weak}
\end{equation}
The corresponding compact and dilute stationary energies therefore
satisfy
\begin{equation}
\begin{aligned}
\mathcal E_{N,\mathrm{eq}}^{\mathrm{comp}}
&\simeq
3k_{B}T_{\beta},
&
\mathcal E_{N,\mathrm{eq}}^{\mathrm{dil}}
&\simeq
\frac{3N}{2}k_{B}T_{\beta},
\\
\frac{
\mathcal E_{N,\mathrm{eq}}^{\mathrm{comp}}
}{
\mathcal E_{N,\mathrm{eq}}^{\mathrm{dil}}
}
&\simeq
\frac{2}{N}.
\end{aligned}
\label{eq:compact-dilute-energy-comparison}
\end{equation}
The physically relevant consequence is the non-extensive scaling of
the stationary mean kinetic energy: in the leading compact closure,
the total stationary energy approaches an $N$-independent value,
whereas the dilute result remains proportional to $N$. The additional
pair friction therefore compensates the linear growth of the
one-particle heating contribution.

Equation~\eqref{eq:Tcomp-physical} should not be applied to $N=2$. For two particles,
the leading compact internal-energy contribution vanishes, and the
correction is controlled instead by the total-momentum variance, as
discussed below Eq.~\eqref{eq:TN_compact_revised}.

The compact result has been derived within a bosonic fixed-$N$
sector. Its direct transfer to quantum-degenerate fermionic matter
requires caution. The leading compact operator identity is independent
of exchange statistics once a configuration satisfying
$r_{ab}\ll r_C$ is assumed. Fermionic antisymmetry, however, produces
an exchange hole for same-spin pairs and introduces a Fermi-pressure
kinetic-energy scale. These effects can modify the finite-$r_C$
crossover and may prevent the low-energy compact closure from being
physically realized. Therefore, Eq.~\eqref{eq:Tcomp-physical} should be regarded as a
bosonic, or effective-bosonic, many-body result and should not be
applied quantitatively to overlapping same-spin fermions without
evaluating the corresponding two-particle reduced density matrix.

\paragraph{Moment hierarchy and non-Gaussian stationary states.}

The closure of the mean-energy balance does not imply the closure of
the full hierarchy of energy moments. In the general fixed-$N$
dynamics, the evolution of the second operator moment
$\mathbb{E}[\langle H_{N}^{2}\rangle_{t}]$ couples to additional
fourth-order single-particle and interparticle momentum correlations.
It therefore cannot, in general, be determined from
$\mathbb{E}[\langle H_{N}\rangle_{t}]$ and
$\mathbb{E}[\langle H_{N}^{2}\rangle_{t}]$ alone without further
assumptions on the state.

This observation is consistent with a recent one-particle
phase-space analysis of the same linear-friction collapse mechanism
\cite{Melo2026}. In that treatment, the equations for the first and
second phase-space moments remain closed, while the first deviations from
Gaussianity appear at fourth order. Finite dissipation generates
enhanced kurtosis and non-Gaussian momentum tails, and the system
approaches a non-equilibrium stationary state rather than a Gibbs
state.

Both $T_{\mathrm{eff}}$ in Eq.~\eqref{eq:Teff-physical} and
$T_{N}^{\mathrm{comp}}$ in Eq.~\eqref{eq:Tcomp-physical} should
therefore be interpreted as parametrizations of stationary
second-moment kinetic-energy scales. Neither quantity, by itself,
establishes thermalization of the full state.

\paragraph{Experimental normalization and bounds.}

To avoid confusion with the stochastic coefficient $\lambda$ defined
in Eq.~\eqref{eq:lambda_definition}, we denote the conventional CSL
collapse rate by $\lambda_{\mathrm{CSL}}$. In the conventions used here,
\begin{equation}
\lambda_{\mathrm{CSL}}
=
\frac{\gamma}
{(4\pi r_{C}^{2})^{3/2}},
\qquad
\gamma
=
(4\pi r_{C}^{2})^{3/2}\lambda_{\mathrm{CSL}}.
\label{eq:gamma-lambda-map}
\end{equation}
Some formulations write the dissipator directly in terms of the mass
density. Their coefficient, also frequently denoted by $\gamma$,
differs from ours by a factor $m_{0}^{-2}$.

For the reference value $r_{C}=10^{-7}\,\mathrm{m}$,
\begin{equation}
\gamma
\simeq
4.45\times10^{-36}\,\mathrm{m^{3}\,s^{-1}}
\left(
\frac{\lambda_{\mathrm{CSL}}}
{10^{-16}\,\mathrm{s^{-1}}}
\right)
\left(
\frac{r_{C}}
{10^{-7}\,\mathrm{m}}
\right)^{3}.
\label{eq:gamma-numerical}
\end{equation}
Thus the GRW and central Adler benchmarks,
$\lambda_{\mathrm{CSL}}=10^{-16}\,\mathrm{s^{-1}}$ and
$10^{-8}\,\mathrm{s^{-1}}$, correspond respectively to
$\gamma\simeq4.45\times10^{-36}\,\mathrm{m^{3}\,s^{-1}}$ and
$4.45\times10^{-28}\,\mathrm{m^{3}\,s^{-1}}$. These values are quoted
only as normalization benchmarks. Experimental constraints are
usually presented directly in the
$(\lambda_{\mathrm{CSL}},r_{C})$ plane, and matter-wave
interferometry is comparatively insensitive to dissipative
corrections over a broad temperature range \cite{Toros2017}.

The linear-friction analysis of Ref.~\cite{DiBartolomeo2024}
excludes the full dCSL parameter space for
\begin{equation}
T_{\beta}\lesssim6\times10^{-9}\,\mathrm{K}.
\label{eq:experimental-Tbeta}
\end{equation}
At $r_{C}=10^{-7}\,\mathrm{m}$, this value corresponds to
$\chi\simeq2.0\times10^{3}$ for a nucleon and therefore lies far
outside the stable one-particle domain. The optomechanical
calculation, however, is based on a harmonic center-of-mass reduction,
for which the relevant mass is the total mechanical mass $M$, rather
than the mass of an individual constituent.

The condition in Eq.~\eqref{eq:chi-stability-3d} can equivalently be
written as
\begin{equation}
m>
m_{\mathrm{crit}}
=
\frac{15\hbar^{2}}
{32r_{C}^{2}k_{B}T_{\beta}},
\end{equation}
or
\begin{equation}
\frac{m_{\mathrm{crit}}}{m_{0}}
\simeq
2.27\times10^{-5}
\left(
\frac{1\,\mathrm{K}}{T_{\beta}}
\right)
\left(
\frac{10^{-7}\,\mathrm{m}}{r_{C}}
\right)^{2}.
\label{eq:critical-mass}
\end{equation}
At $T_{\beta}=6\times10^{-9}\,\mathrm{K}$, one obtains
$m_{\mathrm{crit}}\simeq3.8\times10^{3}\,\mathrm{amu}$. The
nanoparticles and micromagnets employed in the optomechanical bounds
have masses many orders of magnitude above this value, and their
center-of-mass dynamics therefore lies deeply in the $\chi\ll1$
regime.

The distinction between constituent and center-of-mass masses is
essential. In the fixed-$N$ field-theory equations, $m$ denotes the
mass of each bosonic constituent. Replacing it by the total mass of a
rigid body is justified only after an independent center-of-mass
reduction and the inclusion of the corresponding mass form factor.

Finally, Eq.~\eqref{eq:gamma-lambda-map} gives
\begin{equation}
\boxed{
\Gamma_{\mathrm{eff}}^{(1)}
=
2\lambda_{\mathrm{CSL}}
\left(
\frac{m}{m_{0}}
\right)^{2}
\chi
\left(
1-\frac{15}{8}\chi
\right).
}
\label{eq:Gamma-lambda-chi}
\end{equation}
This relation makes explicit the distinct roles of
$\lambda_{\mathrm{CSL}}$ and $\chi$. At fixed $\chi$, decreasing
$\lambda_{\mathrm{CSL}}$ leaves the stationary energy scale unchanged
but increases the relaxation time
$\tau_{\mathrm{rel}}=1/\Gamma_{\mathrm{eff}}^{(1)}$.

For $m=m_{0}$, $r_{C}=10^{-7}\,\mathrm{m}$,
$T_{\beta}=1\,\mathrm{K}$, and the GRW benchmark,
\begin{equation}
\tau_{\mathrm{rel}}
\simeq
1.3\times10^{13}\,\mathrm{yr}
\left(
\frac{10^{-16}\,\mathrm{s^{-1}}}
{\lambda_{\mathrm{CSL}}}
\right).
\label{eq:relaxation-time-benchmark}
\end{equation}
A formally finite stationary energy scale may therefore be
experimentally irrelevant when the relaxation time greatly exceeds
the observation time. The nondissipative limit is correspondingly
singular: taking $\chi\to0$ before the long-time limit recovers the
unbounded standard-CSL heating, whereas any fixed $\chi>0$ formally
produces a finite stationary second moment whenever the relevant
damping coefficient is positive.

\section{Conclusions}
\label{sec:conclusions}

In this work we developed a wave-functional formulation of minimal and dissipative CSL dynamics for a non-relativistic bosonic field. The main advantage of this approach is that the stochastic modification is introduced once at the level of the full Fock-space wave functional, while its consequences in ordinary configuration space are obtained by projection onto fixed particle-number sectors. This provides a systematic way of deriving the one-particle, two-particle, and general \(N\)-particle collapse equations from a single functional dynamics.

The fixed-\(N\) projection makes the amplification mechanism particularly transparent. The collapse operator acts on the total smeared density of the configuration, rather than on isolated particles. As a result, the stochastic and deterministic localization terms naturally acquire collective contributions. In compact configurations, where many particles lie within the same localization length, these contributions add coherently and generate the characteristic many-body amplification of CSL-type models. In dilute configurations, the same formalism shows how the amplification is reduced to a sum over approximately independent correlated blocks.

We also analyzed the probability balance associated with the projected dynamics. In the minimal CSL case, the total norm is preserved by the centered stochastic evolution, but the ordinary local continuity equation is replaced by a stochastic balance equation. This distinction is important for trajectory-based interpretations. In particular, the Bohmian guidance law remains a useful diagnostic tool for the phase dynamics, but exact equivariance is not maintained realization by realization unless the collapse-induced source terms are also encoded in an additional, non-unique probability current.

The dissipative extension considered here further enriches this structure. By adding a current-dependent contribution to the collapse operator, the projected dynamics acquires both real and imaginary modifications. In the polar decomposition, these terms generate dissipative corrections to the probability balance and to the Hamilton-Jacobi equation. In the one-particle sector, the equations reproduce the known balance between collapse-induced heating and friction and generate direct stochastic and deterministic corrections to both density and phase. In the many-particle sector, however, new collective and pair-mixing contributions appear. These terms do not reduce, in general, to a simple sum of one-particle dissipative effects. This shows that dissipation introduces genuinely many-body structures in CSL dynamics.


A central outcome of the analysis is therefore that the functional formulation separates two different aspects of collapse dynamics. On the one hand, it preserves the sectorial structure associated with number-conserving collapse operators: fixed-\(N\) sectors remain closed, and superpositions of different particle-number sectors evolve through stochastic changes of their relative weights rather than through particle creation or annihilation. On the other hand, within each fixed sector, the local dynamics of density, phase, current, and quantum potential can become highly nontrivial, especially in the presence of dissipative current couplings. One of the main new results is the identification of genuinely many-body
dissipative pair terms in the fixed-\(N\) dynamics. In the dilute regime these
terms are suppressed, and the dissipative balance approximately reduces to a
sum of one-particle contributions. In the compact regime, however, the pair
sector produces an additional collective friction and prevents the energy
balance from being generically written as a closed one-particle-like equation.
Within the leading compact closure, this non-extensive behavior can be
made explicit. In three dimensions and for weak dissipation,
$T_N^{\rm comp}\simeq 2T_\beta/N$, so that the stationary total kinetic
energy becomes approximately independent of $N$, rather than growing
linearly as in the dilute regime. This result is conditional on a
negligible total-momentum dispersion and does not imply relaxation of
the full many-body state to a Gibbs distribution.
The framework developed here also suggests several extensions. The dissipative coupling studied in this paper is only one possible choice. Since the wave-functional formalism allows one to define collapse operators directly at the field-theoretic level, other dissipative structures may be explored systematically. For example, one may consider collapse operators involving kinetic-energy density, stress tensors, higher spatial derivatives of the current, or other local bilinear operators. Such generalizations could lead to different effective temperatures, modified energy-balance relations, or new forms of dissipative quantum potentials.

A distinct limitation concerns particle statistics. The compact
non-extensive energy scale derived here follows from a bosonic
fixed-$N$ sector. Although the leading compact operator identity is not altered merely by imposing antisymmetry, the physical realization of that limit can change substantially for fermions. Exchange holes suppress close same-spin configurations, modify the finite-size pair averages entering the crossover away from the strict compact limit, and are accompanied by a Fermi-pressure kinetic-energy scale. The quantity $T_N^{\rm comp}$ should therefore not be transferred directly to quantum-degenerate fermionic systems.

This limitation does not require a fermionic reformulation within the present work. Such a treatment would involve a Grassmann coherent-state  functional representation, antisymmetric sector projections, and a rederivation of the probability, phase, and energy-balance equations.
A useful intermediate extension would be to evaluate the pair term using the two-particle reduced density matrix of a reference Fermi gas, thereby estimating how exchange correlations modify the crossover between the compact and dilute regimes. Conversely, the present bosonic description may remain appropriate for effective bosonic constituents, such as even-even nuclei, bosonic atoms, or localized composite mass units whose exchange is negligible on the scale
resolved by the CSL kernel.


\begin{acknowledgments}
YMPG is supported by a postdoctoral grant from
Fundação Carlos Chagas Filho de Amparo à Pesquisa
do Estado do Rio de Janeiro (FAPERJ), Grant No. E26/200427/2025.
\end{acknowledgments}

\appendix

\section{One-particle dissipative projection}
\label{app:one_particle_dissipative_projection}

In this appendix we derive the one-particle dissipative CSL equation and its
polar form. The calculation fixes the relative factors of $\lambda$, $m$, and
$\kappa$ appearing in Sec.~\ref{subsec:one_particle_consistency}. We assume
throughout that the wave function and the localization profile decrease
sufficiently rapidly for all boundary terms to vanish.

The one-particle component of the Bargmann wave functional is
\begin{equation}
\Psi_1[\bar\alpha,t]
=
\int d^d x\,\psi_1(x,t)\bar\alpha(x),
\label{eq:appA_Psi1}
\end{equation}
with
\begin{equation}
\psi_1(x,t)
=
\left.
\frac{\delta\Psi[\bar\alpha,t]}{\delta\bar\alpha(x)}
\right|_{\bar\alpha=0}.
\label{eq:appA_projection}
\end{equation}
The mass-density part projects according to
\begin{equation}
\left.
\frac{\delta}{\delta\bar\alpha(x)}
\mathcal M_y\Psi_1
\right|_{\bar\alpha=0}
=
mL_y(x)\psi_1(x,t).
\label{eq:appA_mass_projection}
\end{equation}

The functional current operator is
\begin{equation}
\begin{aligned}
\boldsymbol{\mathcal J}_y
={}&
-\frac{i\hbar}{2}
\int d^d u\,L_y(u)
\Bigg[
\bar\alpha(u)\nabla_u
\frac{\delta}{\delta\bar\alpha(u)}
\\
&\hspace{29mm}
-
\bigl(\nabla_u\bar\alpha(u)\bigr)
\frac{\delta}{\delta\bar\alpha(u)}
\Bigg].
\end{aligned}
\label{eq:appA_functional_current}
\end{equation}
Acting on Eq.~\eqref{eq:appA_Psi1} and integrating the second term by
parts gives
\begin{equation}
\begin{aligned}
\boldsymbol{\mathcal J}_y\Psi_1
={}&
\int d^d u\,\bar\alpha(u)
\Bigg[
-i\hbar L_y(u)\nabla_u\psi_1(u,t)
\\
&\hspace{17mm}
-\frac{i\hbar}{2}
\bigl(\nabla_uL_y(u)\bigr)\psi_1(u,t)
\Bigg].
\end{aligned}
\label{eq:appA_current_action}
\end{equation}
Consequently,
\begin{equation}
\boldsymbol{\mathcal J}^{(1)}_y
=
\frac12\left\{L_y(x),-i\hbar\nabla_x\right\}.
\label{eq:appA_current_anticommutator}
\end{equation}
Using $L_y(x)=L(x-y)$ and $\nabla_yL_y(x)=-\nabla_xL_y(x)$, one
finds
\begin{equation}
\nabla_y\cdot\boldsymbol{\mathcal J}^{(1)}_y\psi_1
=
i\hbar
\left[
\nabla L_y(x)\cdot\nabla
+\frac12\nabla^2L_y(x)
\right]\psi_1.
\label{eq:appA_div_current}
\end{equation}
The projected collapse operator is therefore
\begin{equation}
\ell_y^{(1)}
=
mL_y(x)+\kappa A_y,
\qquad
A_y
=
\nabla L_y(x)\cdot\nabla
+\frac12\nabla^2L_y(x),
\label{eq:appA_ell1}
\end{equation}
where $\kappa=\hbar^2\beta/4$. For real $L_y$, integration by parts
shows that
\begin{equation}
A_y^\dagger=-A_y,
\qquad
\ell_y^{(1)\dagger}=mL_y(x)-\kappa A_y.
\label{eq:appA_Ay_antihermitian}
\end{equation}
It follows that the real centering term is
\begin{equation}
r_t(y)
=
\frac12
\left\langle
\ell_y^{(1)\dagger}+\ell_y^{(1)}
\right\rangle_t
=
m\bar N_t(y),
\label{eq:appA_rt}
\end{equation}
where
\begin{equation}
\bar N_t(y)
=
\int d^d x\,\rho_1(x,t)L_y(x),
\qquad
\rho_1=|\psi_1|^2.
\end{equation}

Defining
\begin{equation}
\Delta M_y(x,t)
=
m\bigl[L_y(x)-\bar N_t(y)\bigr],
\label{eq:appA_DeltaM}
\end{equation}
we write the normalized one-particle equation in the compact form
\begin{equation}
d\psi_1
=
\left[-ih_xdt+d\mathcal B_1+\mathcal D_1dt\right]\psi_1,
\label{eq:appA_dpsi_compact}
\end{equation}
where $h_x=-\nabla^2/(2m)$ and
\begin{equation}
d\mathcal B_1
=
\frac{\sqrt\gamma}{m_0}
\int d^d y\,
\bigl(\Delta M_y+\kappa A_y\bigr)dW_t(y).
\label{eq:appA_noise_operator}
\end{equation}
The deterministic non-Hamiltonian operator is
\begin{equation}
\begin{aligned}
\mathcal D_1
={}&
-\frac{\gamma}{2m_0^2}
\int d^d y\,
\Bigl[
\Delta M_y^2
+[mL_y,\kappa A_y]
\\
&\hspace{24mm}
-\kappa^2A_y^2
-2m\kappa\bar N_t(y)A_y
\Bigr].
\end{aligned}
\label{eq:appA_D1_initial}
\end{equation}
The commutator is purely multiplicative,
\begin{equation}
[mL_y,\kappa A_y]
=
-m\kappa|\nabla L_y(x)|^2.
\label{eq:appA_commutator}
\end{equation}

To reduce the $A_y^2$ term, let
\begin{equation}
a_i(y;x)=\partial_iL_y(x),
\qquad
b(y;x)=\nabla^2L_y(x),
\end{equation}
so that $A_y=a_i\partial_i+b/2$. Acting on a test function $\psi$,
\begin{equation}
\begin{aligned}
A_y^2\psi
={}&
a_ia_j\partial_i\partial_j\psi
+
\bigl(a_i\partial_i a_j+ba_j\bigr)\partial_j\psi
\\
&+
\left(
\frac12a_i\partial_i b
+
\frac14b^2
\right)\psi.
\end{aligned}
\label{eq:appA_A2_expanded}
\end{equation}
Translational invariance and the boundary conditions imply
\begin{align}
\int d^d y\,
\bigl(a_i\partial_i a_j+ba_j\bigr)
&=0,
\nonumber\\
\int d^d y\,a_i\partial_i b
&=-\int d^d y\,b^2.
\label{eq:appA_A2_identities}
\end{align}
Therefore,
\begin{equation}
\int d^d y\,A_y^2
=
C_{ij}\partial_i\partial_j
-
\frac14K_L,
\label{eq:appA_A2_reduced}
\end{equation}
where
\begin{align}
C_{ij}
&=
\int d^d y\,
\partial_iL_y(x)\partial_jL_y(x),
\nonumber\\
K_L
&=
\int d^d y\,
\bigl(\nabla^2L_y(x)\bigr)^2.
\label{eq:appA_Cij_KL}
\end{align}
Both contractions are independent of $x$. Substitution into
Eq.~\eqref{eq:appA_D1_initial} gives
\begin{equation}
\begin{aligned}
\mathcal D_1
={}&
-\frac{\gamma}{2m_0^2}
\int d^d y\,
\Bigl[
\Delta M_y^2
-m\kappa|\nabla L_y|^2
\\
&\hspace{15mm}
-2m\kappa\bar N_t(y)A_y
\Bigr]
\\
&\hspace{15mm}+
\frac{\gamma\kappa^2}{2m_0^2}
\left(
C_{ij}\partial_i\partial_j
-
\frac14K_L
\right).
\end{aligned}
\label{eq:appA_D1_final}
\end{equation}
Equations~\eqref{eq:appA_dpsi_compact}, \eqref{eq:appA_noise_operator}, and
\eqref{eq:appA_D1_final} constitute the projected one-particle stochastic
equation used below.

We now derive the polar equations. Write
\begin{equation}
\psi_1=\sqrt{\rho_1}\,e^{iS_1},
\qquad
q_i=\partial_i\ln\sqrt{\rho_1},
\qquad
p_i=\partial_iS_1.
\label{eq:appA_polar_definitions}
\end{equation}
Thus $\partial_i\psi_1/\psi_1=q_i+ip_i$ and
\begin{equation}
\frac{A_y\psi_1}{\psi_1}=u_y+i v_y,
\end{equation}
where
\begin{equation}
u_y=q_i\partial_iL_y+\frac12\nabla^2L_y,
\qquad
v_y=p_i\partial_iL_y.
\label{eq:appA_uv_definitions}
\end{equation}
Introduce the smeared noise
\begin{align}
d\mathcal W_t(x)
&=
\int d^d y\,L_y(x)dW_t(y),
\nonumber\\
\mathbb E[d\mathcal W_t(x)d\mathcal W_t(z)]
&=
g(x-z)dt.
\label{eq:appA_smeared_noise}
\end{align}
The stochastic logarithmic increment is
\begin{equation}
dZ_1^{\rm st}
=
d\Omega_1^{\rm diss}+i\,d\Theta_1^{\rm diss},
\end{equation}
with
\begin{align}
d\Omega_1^{\rm diss}
={}&
d\Omega_1
+
\frac{\sqrt\gamma\kappa}{m_0}
\left[
q_i\partial_i d\mathcal W_t(x)
+
\frac12\nabla^2d\mathcal W_t(x)
\right],
\label{eq:appA_dOmega}
\\
d\Theta_1^{\rm diss}
={}&
\frac{\sqrt\gamma\kappa}{m_0}
 p_i\partial_i d\mathcal W_t(x),
\label{eq:appA_dTheta}
\end{align}
where
\begin{equation}
d\Omega_1
=
\lambda
\int d^d y\,
\bigl[L_y(x)-\bar N_t(y)\bigr]dW_t(y),
\qquad
\lambda=\frac{\sqrt\gamma m}{m_0}.
\label{eq:appA_minimal_increment}
\end{equation}

For the density, It\^o's rule gives
\begin{equation}
d\rho_1
=
\psi_1^*d\psi_1
+
\psi_1d\psi_1^*
+
d\psi_1^*d\psi_1.
\label{eq:appA_Ito_density}
\end{equation}
The terms linear in $\kappa$ in the deterministic part cancel after integration
over $y$, because
\begin{equation}
\int d^d y\,L_y(x)\partial_iL_y(x)
=
\frac12\partial_i g(0)
=
0.
\label{eq:appA_linear_density_cancel}
\end{equation}
For the quadratic contribution one uses
\begin{equation}
\rho_1\int d^d y\,
\left[
\operatorname{Re}\left(
\frac{A_y^2\psi_1}{\psi_1}
\right)
+u_y^2+v_y^2
\right]
=
\frac12C_{ij}\partial_i\partial_j\rho_1.
\label{eq:appA_density_quadratic_identity}
\end{equation}
Combining these identities with the Hamiltonian current yields
\begin{equation}
\begin{aligned}
d\rho_1
={}&
-\partial_i\left[
\frac{\rho_1}{m}
\left(
\partial_iS_1
-
\frac{\lambda^2\kappa^2}{2m}
C_{ij}\partial_j\ln\rho_1
\right)
\right]dt
\\
&+
2\rho_1d\Omega_1
+
\frac{\lambda\kappa}{m}
\partial_i\left[
\rho_1\partial_i d\mathcal W_t(x)
\right].
\end{aligned}
\label{eq:appA_drho}
\end{equation}
Equivalently, the deterministic current-dependent term is the constant-coefficient
diffusion
\begin{equation}
\left(d\rho_1\right)_{\kappa^2}
=
\frac{\lambda^2\kappa^2}{2m^2}
C_{ij}\partial_i\partial_j\rho_1\,dt.
\label{eq:appA_density_diffusion}
\end{equation}

The phase is obtained from the imaginary part of the logarithmic It\^o
increment,
\begin{equation}
dS_1
=
\operatorname{Im}\left[
\frac{d\psi_1}{\psi_1}
-
\frac12
\left(
\frac{d\psi_1}{\psi_1}
\right)^2
\right].
\label{eq:appA_Ito_phase}
\end{equation}
Define
\begin{equation}
\mathcal G_1(x,t)
=
\int d^d z\,\rho_1(z,t)g(z-x).
\label{eq:appA_G1}
\end{equation}
The term linear in $\kappa$ follows from
\begin{equation}
\int d^d y\,
\bigl[2\bar N_t(y)-L_y(x)\bigr]
\partial_iL_y(x)
=
2\partial_i\mathcal G_1(x,t),
\label{eq:appA_phase_linear_identity}
\end{equation}
where the contribution proportional to
$\int d^d y\,L_y\partial_iL_y$ vanishes. The quadratic term is fixed by
\begin{equation}
\int d^d y\,
\left[
\frac12\operatorname{Im}\left(
\frac{A_y^2\psi_1}{\psi_1}
\right)
-u_yv_y
\right]
=
\frac12C_{ij}\partial_i\partial_jS_1.
\label{eq:appA_phase_quadratic_identity}
\end{equation}
Consequently,
\begin{equation}
\begin{aligned}
dS_1
={}&
\left[
-\frac{(\nabla S_1)^2}{2m}
-Q_1
\right]dt
+
\frac{2\lambda^2\kappa}{m}
\partial_iS_1\partial_i\mathcal G_1\,dt
\\
&+
\frac{\lambda^2\kappa^2}{2m^2}
C_{ij}\partial_i\partial_jS_1\,dt
+
\frac{\lambda\kappa}{m}
\partial_iS_1\partial_i d\mathcal W_t(x),
\end{aligned}
\label{eq:appA_dS}
\end{equation}
where
\begin{equation}
Q_1
=
-\frac{1}{2m}
\frac{\nabla^2\sqrt{\rho_1}}{\sqrt{\rho_1}}.
\label{eq:appA_Q1}
\end{equation}
Equations~\eqref{eq:appA_drho} and \eqref{eq:appA_dS} reproduce the
one-particle relations quoted in Sec.~\ref{subsec:one_particle_consistency}.
They also make explicit the qualitative difference from minimal CSL: the
current-dependent collapse operator generates a stochastic probability current,
a direct stochastic phase increment, and deterministic corrections of orders
$\kappa$ and $\kappa^2$.

\section{Fixed-\texorpdfstring{$N$}{N} polar decomposition}
\label{app:fixed_N_polar_decomposition}

In this appendix we derive the fixed-$N$ probability and phase equations quoted in
Sec.~\ref{subsec:fixed_N_collective_pair_terms}. We keep the particle labels explicit
rather than introducing empirical momentum and logarithmic-amplitude fields. This form
shows directly how the current-dependent collapse operator generates a deterministic
pair drift, correlated diffusion in configuration space, and mixed phase-curvature
terms. We assume throughout that the wave function and the localization profile decay
sufficiently rapidly for all boundary terms to vanish.

Starting from Eq.~\eqref{eq:dissipative_fixed_N_revised}, define
\begin{align}
\mathcal M_y^{(N)}
&=m\sum_{a=1}^N L_y(x_a),
&
\mathcal A_y^{(N)}
&=\sum_{a=1}^N A_{a,y},
\nonumber\\
B_y
&=\mathcal M_y^{(N)}-r_y.
\label{eq:appB_MAB_definitions}
\end{align}
The one-particle differential operator entering the sum is
\begin{equation}
\begin{aligned}
A_{a,y}
={}&
\nabla_aL_y(x_a)\cdot\nabla_a
\\
&+
\frac12\nabla_a^2L_y(x_a).
\end{aligned}
\label{eq:appB_Aay_definition}
\end{equation}
Then
\begin{equation}
\ell_y^{(N)}-r_y
=
B_y+\kappa\mathcal A_y^{(N)},
\end{equation}
and, since $\mathcal A_y^{(N)\dagger}=-\mathcal A_y^{(N)}$,
\begin{equation}
\begin{aligned}
&\ell_y^{(N)\dagger}\ell_y^{(N)}+r_y^2-2r_y\ell_y^{(N)}
\\
&\qquad =
B_y^2
+\kappa[\mathcal M_y^{(N)},\mathcal A_y^{(N)}]
-\kappa^2\bigl(\mathcal A_y^{(N)}\bigr)^2
-2r_y\kappa\mathcal A_y^{(N)}.
\end{aligned}
\label{eq:appB_operator_expansion}
\end{equation}
Derivatives with respect to $x_a$ do not act on $L_y(x_b)$ for $a\neq b$.
Consequently, the commutator is diagonal in the particle labels:
\begin{equation}
[\mathcal M_y^{(N)},\mathcal A_y^{(N)}]
=
-m\sum_{a=1}^N
\left|\nabla_aL_y(x_a)\right|^2.
\label{eq:appB_commutator}
\end{equation}

Write
\begin{equation}
\psi_N(X_N,t)=\sqrt{\rho_N(X_N,t)}\,e^{iS_N(X_N,t)},
\end{equation}
and define
\begin{equation}
q_{a,i}=\partial_{a i}\ln\sqrt{\rho_N},
\qquad
p_{a,i}=\partial_{a i}S_N.
\label{eq:appB_pq_definitions}
\end{equation}
It is also useful to introduce
\begin{equation}
l_{a i}(y)\equiv\partial_{a i}L_y(x_a),
\qquad
d_a(y)\equiv\nabla_a^2L_y(x_a).
\label{eq:appB_ld_definitions}
\end{equation}
The total current operator then acts according to
\begin{equation}
\frac{\mathcal A_y^{(N)}\psi_N}{\psi_N}
=u_y+i v_y,
\label{eq:appB_uv_action}
\end{equation}
where
\begin{align}
u_y
&=
\sum_{a=1}^N
\left[q_{a,i}l_{a i}(y)+\frac12d_a(y)\right],
\nonumber\\
v_y
&=
\sum_{a=1}^N p_{a,i}l_{a i}(y).
\label{eq:appB_uv_definitions}
\end{align}
Repeated spatial indices are summed. The kernel contractions required below are
\begin{align}
\int d^d y\,L_y(x_b)\partial_{a i}L_y(x_a)
&=
\partial_{a i}g(x_a-x_b),
\label{eq:appB_kernel_LdL}
\\
C_{ij}(x_a,x_b)
&\equiv
\int d^d y\,
\partial_{a i}L_y(x_a)\partial_{b j}L_y(x_b)
\nonumber\\
&=
-\partial_i\partial_jg(x_a-x_b).
\label{eq:appB_kernel_Cij}
\end{align}
For coincident arguments, translational invariance implies
$\partial_{a i}g(0)=0$, while $C_{ij}(x_a,x_a)$ is independent of $x_a$.

\subsection{Probability equation}

It\^o's rule gives
\begin{equation}
d\rho_N
=
\psi_N^*d\psi_N
+\psi_Nd\psi_N^*
+d\psi_N^*d\psi_N.
\label{eq:appB_Ito_density}
\end{equation}
The real stochastic increment is
\begin{equation}
2\rho_N\frac{\sqrt\gamma}{m_0}
\int d^d y\,[B_y+\kappa u_y]dW_t(y).
\label{eq:appB_density_noise_initial}
\end{equation}
The term proportional to $B_y$ is the minimal-CSL contribution
$2\rho_Nd\Omega_N$. Introducing the smeared noise
\begin{equation}
d\mathcal W_t(x)
=
\int d^d y\,L_y(x)dW_t(y),
\end{equation}
the current-dependent stochastic term becomes
\begin{equation}
2\rho_N\frac{\sqrt\gamma\kappa}{m_0}
\int d^d y\,u_y dW_t(y)
=
\frac{\lambda\kappa}{m}
\sum_{a=1}^N
\nabla_a\cdot
\left[\rho_N\nabla_a d\mathcal W_t(x_a)\right].
\label{eq:appB_stochastic_density_current}
\end{equation}

We next combine the deterministic operator in
Eq.~\eqref{eq:appB_operator_expansion} with the quadratic It\^o term. At order
$\kappa$, the terms proportional to the centering combine with those containing
$B_y$ so that only $\mathcal M_y^{(N)}=B_y+r_y$ remains. One obtains the exact
identity
\begin{equation}
\begin{aligned}
&\rho_N
\left[
-[\mathcal M_y^{(N)},\mathcal A_y^{(N)}]
+2\mathcal M_y^{(N)}u_y
\right]
\\
&\qquad =
 m\sum_{a,b=1}^N
\nabla_a\cdot
\left[
\rho_N L_y(x_b)\nabla_aL_y(x_a)
\right].
\end{aligned}
\label{eq:appB_linear_density_identity}
\end{equation}
After multiplication by the factor $\gamma\kappa/m_0^2$ and integration over
the collapse center, Eq.~\eqref{eq:appB_kernel_LdL} gives
\begin{equation}
\frac{\lambda^2\kappa}{m}
\sum_{a,b=1}^N
\nabla_a\cdot
\left[
\rho_N\nabla_a g(x_a-x_b)
\right]dt.
\label{eq:appB_pair_drift_all_labels}
\end{equation}
The diagonal terms vanish because $\nabla g(0)=0$, leaving only $a\neq b$.

At order $\kappa^2$, it is useful to regard the particle and spatial labels as
a single composite index $\mu=(a,i)$. For a fixed collapse center, the
current-dependent density drift obeys
\begin{equation}
\begin{aligned}
&\rho_N
\left[
\operatorname{Re}\left(
\frac{(\mathcal A_y^{(N)})^2\psi_N}{\psi_N}
\right)
+u_y^2+v_y^2
\right]
\\
&\qquad =
\frac12\partial_\mu
\left[
l_\mu(y)\partial_\nu
\bigl(l_\nu(y)\rho_N\bigr)
\right].
\end{aligned}
\label{eq:appB_quadratic_density_identity}
\end{equation}
This expression is not a double divergence of
$\rho_Nl_\mu l_\nu$ before the integration over $y$. After that integration,
however, translational invariance implies
\begin{equation}
\sum_\nu\int d^d y\,
l_\mu(y)\partial_\nu l_\nu(y)
=
\sum_\nu\partial_\nu C_{\mu\nu},
\label{eq:appB_C_divergence_identity}
\end{equation}
where $C_{\mu\nu}\equiv C_{ij}(x_a,x_b)$. The additional same-particle term
vanishes as a derivative of the translation-invariant coincident contraction. Consequently,
\begin{equation}
\begin{aligned}
&\int d^d y\,
\rho_N
\left[
\operatorname{Re}\left(
\frac{(\mathcal A_y^{(N)})^2\psi_N}{\psi_N}
\right)
+u_y^2+v_y^2
\right]
\\
&\qquad =
\frac12
\sum_{a,b=1}^N
\partial_{a i}\partial_{b j}
\left[
\rho_N C_{ij}(x_a,x_b)
\right].
\end{aligned}
\label{eq:appB_quadratic_density_integrated}
\end{equation}
Using $\gamma/m_0^2=\lambda^2/m^2$, the quadratic contribution is therefore
\begin{equation}
\frac{\lambda^2\kappa^2}{2m^2}
\sum_{a,b=1}^N
\partial_{a i}\partial_{b j}
\left[
\rho_N C_{ij}(x_a,x_b)
\right]dt.
\label{eq:appB_correlated_diffusion}
\end{equation}
The matrix with composite indices $(a,i)$ and $(b,j)$ is positive semidefinite,
since it is the Gram matrix of the functions $\partial_{a i}L_y(x_a)$.

Combining the Hamiltonian current with
Eqs.~\eqref{eq:appB_stochastic_density_current},
\eqref{eq:appB_pair_drift_all_labels}, and
\eqref{eq:appB_correlated_diffusion}, one obtains
\begin{equation}
\begin{aligned}
d\rho_N
&+
\sum_{a=1}^N
\nabla_a\cdot
\left(
\frac{\rho_N\nabla_aS_N}{m}
\right)dt
=
2\rho_Nd\Omega_N
\\
&+
\frac{\lambda\kappa}{m}
\sum_{a=1}^N
\nabla_a\cdot
\left[
\rho_N\nabla_a d\mathcal W_t(x_a)
\right]
\\
&+
\frac{\lambda^2\kappa}{m}
\sum_{a\neq b}
\nabla_a\cdot
\left[
\rho_N\nabla_a g(x_a-x_b)
\right]dt
\\
&+
\frac{\lambda^2\kappa^2}{2m^2}
\sum_{a,b=1}^N
\partial_{a i}\partial_{b j}
\left[
\rho_N C_{ij}(x_a,x_b)
\right]dt.
\end{aligned}
\label{eq:appB_drhoN_final}
\end{equation}
Every current-dependent term is in divergence form. Hence it integrates to a
boundary term, while the centered minimal-CSL increment integrates to zero. The
normalization of the fixed sector is therefore preserved for each noise realization.

\subsection{Phase equation}

The phase follows from the imaginary part of the logarithmic It\^o increment. With
$c\equiv\sqrt\gamma/m_0$, define
\begin{equation}
\begin{aligned}
\mathcal D_N
={}&
-\frac{c^2}{2}
\int d^d y\,
\Bigl[
B_y^2
+\kappa[\mathcal M_y^{(N)},\mathcal A_y^{(N)}]
\\
&\hspace{14mm}
-\kappa^2(\mathcal A_y^{(N)})^2
-2r_y\kappa\mathcal A_y^{(N)}
\Bigr].
\end{aligned}
\label{eq:appB_DN_definition}
\end{equation}
Define the stochastic logarithmic coefficient
\begin{equation}
\zeta_y
=
B_y+\kappa(u_y+i v_y).
\label{eq:appB_zeta_definition}
\end{equation}
The logarithmic It\^o formula then gives
\begin{equation}
\begin{aligned}
dS_N
={}&
\operatorname{Im}\Bigg[
-i\frac{H_N\psi_N}{\psi_N}dt
+c\int d^d y\,\zeta_y dW_t(y)
\\
&+
\frac{\mathcal D_N\psi_N}{\psi_N}dt
-
\frac{c^2}{2}
\int d^d y\,\zeta_y^2dt
\Bigg].
\end{aligned}
\label{eq:appB_Ito_phase}
\end{equation}
The Hamiltonian term gives
\begin{equation}
-\left[
\sum_{a=1}^N\frac{p_a^2}{2m}+Q_N
\right]dt,\qquad
Q_N
=
-\sum_{a=1}^N
\frac{1}{2m}
\frac{\nabla_a^2\sqrt{\rho_N}}{\sqrt{\rho_N}}.
\label{eq:appB_Hamiltonian_phase}
\end{equation}
The stochastic phase increment is
\begin{equation}
\frac{\lambda\kappa}{m}
\sum_{a=1}^N
p_a\cdot\nabla_a d\mathcal W_t(x_a).
\label{eq:appB_stochastic_phase}
\end{equation}

At order $\kappa$, the deterministic terms combine into
\begin{equation}
c^2\kappa
\int d^d y\,
\left[2r_y-\mathcal M_y^{(N)}\right]v_y.
\label{eq:appB_linear_phase_initial}
\end{equation}
Since
\begin{equation}
r_y=mN\bar\nu_y(t),
\qquad
\mathcal M_y^{(N)}=mN\nu_y[X_N],
\end{equation}
we define
\begin{equation}
\mathcal G_N(x;X_N,t)
=
\int d^d z\,
\left[2\bar\nu(z,t)-\nu_{X_N}(z)\right]g(z-x).
\label{eq:appB_GN_definition}
\end{equation}
The kernel identity
\begin{equation}
\int d^d y\,
\left[2\bar\nu_y(t)-\nu_y[X_N]\right]
\nabla_aL_y(x_a)
=
\nabla\mathcal G_N(x_a;X_N,t)
\label{eq:appB_GN_identity}
\end{equation}
then gives
\begin{equation}
c^2\kappa
\int d^d y\,
\left[2r_y-\mathcal M_y^{(N)}\right]v_y
=
\frac{\eta_N}{m}
\sum_{a=1}^N
p_a\cdot\nabla\mathcal G_N(x_a),
\label{eq:appB_linear_phase_final}
\end{equation}
where
\begin{equation}
\eta_N
=
\frac{\gamma\kappa m^2N}{m_0^2}
=
\lambda^2\kappa N.
\label{eq:appB_etaN}
\end{equation}

The quadratic phase correction is
\begin{equation}
c^2\kappa^2
\int d^d y\,
\left[
\frac12
\operatorname{Im}\left(
\frac{(\mathcal A_y^{(N)})^2\psi_N}{\psi_N}
\right)
-u_yv_y
\right].
\label{eq:appB_quadratic_phase_initial}
\end{equation}
Using
\begin{equation}
\frac{(\mathcal A_y^{(N)})^2\psi_N}{\psi_N}
=
\sum_{a=1}^N l_{a i}(y)\partial_{a i}(u_y+i v_y)
+(u_y+i v_y)^2,
\label{eq:appB_A2_polar_identity}
\end{equation}
one finds
\begin{equation}
\frac12
\operatorname{Im}\left(
\frac{(\mathcal A_y^{(N)})^2\psi_N}{\psi_N}
\right)
-u_yv_y
=
\frac12
\sum_{a=1}^N l_{a i}(y)\partial_{a i}v_y.
\label{eq:appB_quadratic_phase_reduction}
\end{equation}
On expanding $v_y$, a term containing
$l_{a i}\partial_{a i}l_{b j}$ appears. It is nonzero only for $a=b$, and its
integral over the collapse center vanishes:
\begin{equation}
\int d^d y\,
\partial_iL_y(x)\partial_i\partial_jL_y(x)
=
\frac12\partial_j
\int d^d y\,|\nabla L_y(x)|^2
=0.
\label{eq:appB_phase_kernel_derivative_cancel}
\end{equation}
Therefore
\begin{equation}
\begin{aligned}
&\int d^d y\,
\left[
\frac12
\operatorname{Im}\left(
\frac{(\mathcal A_y^{(N)})^2\psi_N}{\psi_N}
\right)
-u_yv_y
\right]
\\
&\qquad =
\frac12
\sum_{a,b=1}^N
C_{ij}(x_a,x_b)
\partial_{a i}\partial_{b j}S_N.
\end{aligned}
\label{eq:appB_quadratic_phase_final}
\end{equation}

Combining the preceding results gives the full phase equation
\begin{equation}
\begin{aligned}
dS_N
={}&
\Bigg[
-\sum_{a=1}^N\frac{p_a^2}{2m}
-Q_N
+
\frac{\eta_N}{m}
\sum_{a=1}^N
p_a\cdot\nabla\mathcal G_N(x_a)
\\
&\hspace{8mm}
+
\frac{\gamma\kappa^2}{2m_0^2}
\sum_{a,b=1}^N
C_{ij}(x_a,x_b)
\partial_{a i}\partial_{b j}S_N
\Bigg]dt
\\
&\hspace{8mm}+
\frac{\lambda\kappa}{m}
\sum_{a=1}^N
p_a\cdot\nabla_a d\mathcal W_t(x_a).
\end{aligned}
\label{eq:appB_dSN_final}
\end{equation}
This is Eq.~\eqref{eq:dSN_full_revised} of the main text.

For the algebraic completion of squares, define
\begin{equation}
\Pi_{a i}^{(N)}
=
p_{a,i}
-
\eta_N\nabla_i\mathcal G_N(x_a;X_N,t).
\label{eq:appB_PiN_definition}
\end{equation}
Then
\begin{equation}
\begin{aligned}
-&\sum_{a=1}^N\frac{p_a^2}{2m}
+
\frac{\eta_N}{m}
\sum_{a=1}^N
p_a\cdot\nabla\mathcal G_N(x_a)
\\
&\hspace{4mm}=
-\sum_{a=1}^N
\frac{(\Pi_a^{(N)})^2}{2m}
+
\frac{\eta_N^2}{2m}
\sum_{a=1}^N
\left|\nabla\mathcal G_N(x_a)\right|^2.
\end{aligned}
\label{eq:appB_completion_square}
\end{equation}
The off-diagonal part of the phase-curvature term is
\begin{equation}
\mathcal R_{\rm mix}^{(N)}
=
\frac{\gamma\kappa^2}{2m_0^2}
\sum_{a\neq b}
C_{ij}(x_a,x_b)
\partial_{a i}\partial_{b j}S_N.
\label{eq:appB_Rmix}
\end{equation}
It is absent for $N=1$ and is suppressed when the particle separations are much
larger than $r_C$. Compactness alone does not make it nonzero: the phase must also
contain mixed curvatures. In particular,
$\mathcal R_{\rm mix}^{(N)}=0$ for an additive phase
$S_N(X_N)=\sum_aS_a(x_a)$. The quantities $\Pi_a^{(N)}$ are therefore only
shifted phase gradients generated by the completion of squares; they do not define
the probability-transport current in Eq.~\eqref{eq:appB_drhoN_final}.

As a consistency check, for $N=1$ the pair drift vanishes,
$C_{ij}(x_1,x_1)=C_{ij}$ is constant, and
\begin{equation}
\mathcal G_N(x_1;X_1,t)
=
2\mathcal G_1(x_1,t)-g(0).
\end{equation}
Hence Eqs.~\eqref{eq:appB_drhoN_final} and \eqref{eq:appB_dSN_final}
reduce exactly to Eqs.~\eqref{eq:appA_drho} and \eqref{eq:appA_dS}.
\section{Energy-balance commutators and Gaussian coefficients}
\label{app:energy_balance_commutators}

In this appendix we derive the energy balances used in
Secs.~\ref{subsec:one_particle_consistency} and~\ref{subsec:energy_balance_revised}. We first evaluate the one-particle
coefficients and then separate the diagonal and off-diagonal contractions in a
fixed-$N$ sector. Throughout the appendix, spatial boundary terms are assumed to
vanish and the kernel is translationally invariant.

\subsection{General energy identity}

Consider a normalized It\^o equation with a possibly non-Hermitian collapse
operator $\ell_y$ and real centering $r_t(y)$,
\begin{equation}
\begin{aligned}
d|\psi_t\rangle
={}&
\Bigg[
-iHdt
+
\frac{\sqrt\gamma}{m_0}
\int d^d y\,\bigl(\ell_y-r_t(y)\bigr)dW_t(y)
\\
&-
\frac{\gamma}{2m_0^2}
\int d^d y\,
\left(
\ell_y^\dagger\ell_y+r_t^2(y)-2r_t(y)\ell_y
\right)dt
\Bigg]|\psi_t\rangle.
\end{aligned}
\label{eq:appC_general_SSE}
\end{equation}
For $\mathcal E(t)=\langle H\rangle_t$, It\^o's rule gives
\begin{equation}
\begin{aligned}
d\mathcal E
={}&
d\mathcal E^{\rm st}
+
\frac{\gamma}{2m_0^2}
\int d^d y\,
\Big\langle
\ell_y^\dagger[H,\ell_y]
\\
&\hspace{10mm}
+[\ell_y^\dagger,H]\ell_y
\Big\rangle_tdt,
\end{aligned}
\label{eq:appC_general_energy_identity}
\end{equation}
where
\begin{equation}
\begin{aligned}
d\mathcal E^{\rm st}
={}&
\frac{\sqrt\gamma}{m_0}
\int d^d y\,
\Big[
\langle\ell_y^\dagger H+H\ell_y\rangle_t
\\
&\hspace{10mm}
-2r_t(y)\mathcal E(t)
\Big]dW_t(y).
\end{aligned}
\label{eq:appC_energy_noise}
\end{equation}
Hence $\mathbb E[d\mathcal E^{\rm st}]=0$. Notice that the centering cancels
completely from the deterministic energy drift.

\subsection{One-particle balance}

For one particle,
\begin{equation}
\ell_y=mL_y+\kappa A_y,
\qquad
A_y=\partial_iL_y\,\partial_i+\frac12\nabla^2L_y,
\label{eq:appC_ell_one}
\end{equation}
with $A_y^\dagger=-A_y$ and $h=-\nabla^2/(2m)$. Expanding the operator in
Eq.~\eqref{eq:appC_general_energy_identity} gives
\begin{equation}
\begin{aligned}
&\ell_y^\dagger[h,\ell_y]+[\ell_y^\dagger,h]\ell_y
\\
&\quad =
[mL_y,[h,mL_y]]
\\
&\qquad
+\kappa m\left(
\{L_y,[h,A_y]\}-\{A_y,[h,L_y]\}
\right)
\\
&\qquad
-\kappa^2[A_y,[h,A_y]].
\end{aligned}
\label{eq:appC_one_particle_expansion}
\end{equation}
The identity
\begin{equation}
[h,L_y]=-\frac{1}{m}A_y
\label{eq:appC_hL_identity}
\end{equation}
immediately yields the minimal-CSL term
\begin{equation}
[mL_y,[h,mL_y]]=m|\nabla L_y|^2.
\label{eq:appC_minimal_term}
\end{equation}
The remaining terms are most transparently evaluated in momentum space.

We use
\begin{equation}
L_y(x)
=
\int\frac{d^d k}{(2\pi)^d}\,
\widetilde L(k)e^{ik\cdot(x-y)},\qquad
|\widetilde L(k)|^2=\widetilde g(k),
\label{eq:appC_L_fourier}
\end{equation}
and write $\widehat{\boldsymbol p}=-i\nabla$. Acting on a momentum eigenstate,
$\ell_y$ transfers momentum $k$ with amplitude
\begin{equation}
\widetilde L(k)e^{-ik\cdot y}
\left[
 m-\kappa\left(k\cdot p+\frac{k^2}{2}\right)
\right].
\label{eq:appC_jump_amplitude}
\end{equation}
The corresponding kinetic-energy change is
\begin{equation}
\frac{(p+k)^2-p^2}{2m}
=
\frac{1}{m}
\left(k\cdot p+\frac{k^2}{2}\right).
\label{eq:appC_energy_transfer}
\end{equation}
Inserting momentum resolutions in Eq.~\eqref{eq:appC_general_energy_identity}
therefore gives the exact deterministic drift
\begin{equation}
\begin{aligned}
\left(\frac{d\mathcal E_1}{dt}\right)_{\rm det}
={}&
\frac{\gamma}{m_0^2}
\int\frac{d^d k}{(2\pi)^d}\,
|\widetilde L(k)|^2
\\
&\times
\left\langle
\left(m-\kappa q_k\right)^2\frac{q_k}{m}
\right\rangle_t,
\end{aligned}
\label{eq:appC_momentum_energy_drift}
\end{equation}
where
\begin{equation}
q_k=k\cdot\widehat{\boldsymbol p}+\frac{k^2}{2}.
\end{equation}
Equation~\eqref{eq:appC_momentum_energy_drift} is already quadratic in $\kappa$;
there are no omitted higher-order terms.

For an isotropic profile, define
\begin{align}
I_2
&=
\int\frac{d^d k}{(2\pi)^d}\,
k^2|\widetilde L(k)|^2,
\nonumber\\
I_4
&=
\int\frac{d^d k}{(2\pi)^d}\,
k^4|\widetilde L(k)|^2,
\nonumber\\
I_6
&=
\int\frac{d^d k}{(2\pi)^d}\,
k^6|\widetilde L(k)|^2.
\label{eq:appC_kernel_moments}
\end{align}
The angular averages satisfy
\begin{align}
\int\frac{d^d k}{(2\pi)^d}\,
k_i k_j|\widetilde L(k)|^2
&=
\frac{I_2}{d}\delta_{ij},
\nonumber\\
\int\frac{d^d k}{(2\pi)^d}\,
k^2k_i k_j|\widetilde L(k)|^2
&=
\frac{I_4}{d}\delta_{ij}.
\label{eq:appC_isotropic_averages}
\end{align}
Expanding the integrand in Eq.~\eqref{eq:appC_momentum_energy_drift} and using
$\langle\widehat{\boldsymbol p}^{\,2}\rangle_t=2m\mathcal E_1(t)$ gives
\begin{equation}
\begin{aligned}
\left(\frac{d\mathcal E_1}{dt}\right)_{\rm det}
={}&
\frac{\gamma}{m_0^2}
\left(
\frac{m}{2}I_2
-
\frac{\kappa}{2}I_4
+
\frac{\kappa^2}{8m}I_6
\right)
\\
&-
\frac{\gamma}{m_0^2}
\left(
\frac{4\kappa m}{d}I_2
-
\frac{3\kappa^2}{d}I_4
\right)\mathcal E_1(t).
\end{aligned}
\label{eq:appC_one_particle_drift_moments}
\end{equation}
Equivalently, introduce the position-space contractions
\begin{align}
C_{ij}(x,x)
&=
\int d^d y\,\partial_iL_y(x)\partial_jL_y(x)
=C_d\delta_{ij},
\nonumber\\
C_{ii}
&=dC_d=I_2,
\nonumber\\
K_L
&=
\int d^d y\,\bigl(\nabla^2L_y(x)\bigr)^2=I_4,
\nonumber\\
S_6
&=I_6.
\label{eq:appC_position_contractions}
\end{align}
The ensemble-averaged balance then takes the form
\begin{equation}
\frac{d}{dt}\mathbb E[\mathcal E_1]
=
P_{\rm eff}^{(1)}
-
\Gamma_{\rm eff}^{(1)}\mathbb E[\mathcal E_1],
\label{eq:appC_one_particle_balance}
\end{equation}
with
\begin{equation}
P_{\rm eff}^{(1)}
=
\frac{\gamma m}{2m_0^2}C_{ii}
-
\frac{\gamma\kappa}{2m_0^2}K_L
+
\frac{\gamma\kappa^2}{8m_0^2m}S_6,
\label{eq:appC_Peff_general}
\end{equation}
and
\begin{equation}
\Gamma_{\rm eff}^{(1)}
=
\frac{4\gamma\kappa m}{m_0^2}C_d
-
\frac{3\gamma\kappa^2}{d m_0^2}K_L.
\label{eq:appC_Gammaeff_general}
\end{equation}
The first term of Eq.~\eqref{eq:appC_Peff_general} is the standard CSL heating.
The term linear in $\kappa$ produces both friction and a negative correction to the
state-independent heating, while the quadratic term partially compensates the friction
and adds a higher-derivative heating contribution.

\subsection{Gaussian coefficients and effective temperature}

For the Gaussian kernel used in the main text,
\begin{equation}
g(r)
=
\frac{1}{(4\pi r_C^2)^{d/2}}
\exp\left(-\frac{r^2}{4r_C^2}\right),
\qquad
\widetilde g(k)=e^{-r_C^2k^2},
\label{eq:appC_gaussian_kernel}
\end{equation}
one obtains
\begin{equation}
C_d
=
\frac{1}{2^{d+1}\pi^{d/2}r_C^{d+2}},
\qquad
C_{ii}=dC_d,
\label{eq:appC_Cd_gaussian}
\end{equation}
\begin{equation}
K_L
=
\frac{d(d+2)}{2^{d+2}\pi^{d/2}r_C^{d+4}},
\label{eq:appC_KL_gaussian}
\end{equation}
and
\begin{equation}
S_6
=
\frac{d(d+2)(d+4)}
{2^{d+3}\pi^{d/2}r_C^{d+6}}.
\label{eq:appC_S6_gaussian}
\end{equation}
Using $\chi=\kappa/(mr_C^2)$, Eqs.~\eqref{eq:appC_Peff_general} and
\eqref{eq:appC_Gammaeff_general} become
\begin{equation}
P_{\rm eff}^{(1)}
=
\frac{\gamma m C_d}{m_0^2}
\left[
\frac{d}{2}
-
\frac{d(d+2)}{4}\chi
+
\frac{d(d+2)(d+4)}{32}\chi^2
\right],
\label{eq:appC_Peff_gaussian}
\end{equation}
and
\begin{equation}
\Gamma_{\rm eff}^{(1)}
=
\frac{\gamma\kappa m C_d}{m_0^2}
\left[
4-
\frac{3(d+2)}{2}\chi
\right].
\label{eq:appC_Gammaeff_gaussian}
\end{equation}
The damping coefficient is positive for
\begin{equation}
0<\chi<\frac{8}{3(d+2)}.
\label{eq:appC_positive_damping}
\end{equation}
The polynomial in Eq.~\eqref{eq:appC_Peff_gaussian} is positive for all real
$\chi$, so within the range \eqref{eq:appC_positive_damping} the stationary energy
is positive and finite.

Defining
\begin{equation}
\mathcal E_{1,{\rm eq}}
=
\frac{P_{\rm eff}^{(1)}}{\Gamma_{\rm eff}^{(1)}}
=
\frac{d}{2}k_BT_{\rm eff}^{(d)},
\end{equation}
one finds
\begin{equation}
k_BT_{\rm eff}^{(d)}
=
\frac{1}{4\kappa}
\frac{
1-\dfrac{d+2}{2}\chi
+\dfrac{(d+2)(d+4)}{16}\chi^2
}{
1-\dfrac{3(d+2)}{8}\chi
}.
\label{eq:appC_Teff}
\end{equation}
For weak dissipation,
\begin{equation}
k_BT_{\rm eff}^{(d)}
=
\frac{1}{4\kappa}
\left[
1-\frac{d+2}{8}\chi+\mathcal O(\chi^2)
\right].
\label{eq:appC_Teff_weak}
\end{equation}
As emphasized in the main text, this temperature parametrizes the stationary kinetic
energy and does not by itself imply a Gibbs stationary state.

\subsection{Fixed-\texorpdfstring{$N$}{N} decomposition}

For a fixed particle-number sector, define
\begin{align}
\mathcal M_y^{(N)}
&=
m\sum_{a=1}^N L_y(x_a),
&
\mathcal A_y^{(N)}
&=
\sum_{a=1}^N A_{a,y},
\nonumber\\
\ell_y^{(N)}
&=
\mathcal M_y^{(N)}+\kappa\mathcal A_y^{(N)},
&
H_N
&=
\sum_{a=1}^N h_a.
\label{eq:appC_fixedN_definitions}
\end{align}
The deterministic operator entering the energy balance is
\begin{equation}
\begin{aligned}
\mathcal F_y^{(N)}
={}&
\ell_y^{(N)\dagger}[H_N,\ell_y^{(N)}]
+[\ell_y^{(N)\dagger},H_N]\ell_y^{(N)}
\\
={}&
[\mathcal M_y^{(N)},[H_N,\mathcal M_y^{(N)}]]
\\
&+
\kappa\{\mathcal M_y^{(N)},[H_N,\mathcal A_y^{(N)}]\}
\\
&-
\kappa\{\mathcal A_y^{(N)},[H_N,\mathcal M_y^{(N)}]\}
\\
&-
\kappa^2
[\mathcal A_y^{(N)},[H_N,\mathcal A_y^{(N)}]].
\end{aligned}
\label{eq:appC_fixedN_operator_expansion}
\end{equation}
Since operators carrying different particle labels commute,
\begin{equation}
[H_N,\mathcal M_y^{(N)}]
=-\mathcal A_y^{(N)},
\label{eq:appC_HM_minus_A}
\end{equation}
and the terms of order $\kappa^0$ and $\kappa^2$ in
Eq.~\eqref{eq:appC_fixedN_operator_expansion} contain only equal-label
contractions. Their sum gives $N$ copies of the state-independent heating and a
friction term proportional to the total kinetic energy $\mathcal E_N$, namely
\begin{equation}
NP_{\rm eff}^{(1)}
-
\Gamma_{\rm eff}^{(1)}\mathcal E_N.
\label{eq:appC_diagonal_fixedN_balance}
\end{equation}
Only the term linear in $\kappa$ contains off-diagonal contractions.

For an unordered pair $a<b$, the off-diagonal part of
Eq.~\eqref{eq:appC_fixedN_operator_expansion} is
\begin{equation}
\begin{aligned}
\mathcal F_{ab}^{\rm off}(y)
=2\kappa\Big[
&mL_y(x_a)[h_b,A_{b,y}]
+mL_y(x_b)[h_a,A_{a,y}]
\\
&+2A_{a,y}A_{b,y}
\Big].
\end{aligned}
\label{eq:appC_Fab_off_initial}
\end{equation}
A direct integration by parts, together with translational invariance, gives
\begin{equation}
\int d^d y\,\mathcal F_{ab}^{\rm off}(y)
=
2\kappa C_{ij}(x_a,x_b)D_{ab,i}D_{ab,j},
\label{eq:appC_Fab_integrated_C}
\end{equation}
where
\begin{align}
C_{ij}(x_a,x_b)
&=
\int d^d y\,
\partial_{a i}L_y(x_a)\partial_{b j}L_y(x_b)
\nonumber\\
&=
-\partial_i\partial_jg(r_{ab}),
\nonumber\\
D_{ab}
&=
\nabla_a+\nabla_b,
\qquad
r_{ab}=x_a-x_b.
\label{eq:appC_Cij_Dab}
\end{align}
Since $D_{ab}g(r_{ab})=0$, the kernel commutes with $D_{ab}$. Introducing
\begin{equation}
\widehat{\boldsymbol P}_{ab}
=-iD_{ab}
=
\widehat{\boldsymbol p}_a+\widehat{\boldsymbol p}_b,
\qquad
g_{ij}(r)=\partial_i\partial_jg(r),
\end{equation}
Eq.~\eqref{eq:appC_Fab_integrated_C} is equivalently
\begin{equation}
\int d^d y\,\mathcal F_{ab}^{\rm off}(y)
=
2\kappa g_{ij}(r_{ab})
\widehat P_{ab,i}\widehat P_{ab,j}.
\label{eq:appC_Fab_integrated_P}
\end{equation}
Substitution into Eq.~\eqref{eq:appC_general_energy_identity} yields
\begin{equation}
\frac{d}{dt}\mathbb E[\mathcal E_N]
=
NP_{\rm eff}^{(1)}
-
\Gamma_{\rm eff}^{(1)}\mathbb E[\mathcal E_N]
+
\mathbb E[\mathcal C_N],
\label{eq:appC_fixedN_energy_balance}
\end{equation}
with
\begin{equation}
\mathcal C_N
=
\frac{\gamma\kappa}{m_0^2}
\sum_{a<b}
\left\langle
 g_{ij}(r_{ab})
 \widehat P_{ab,i}\widehat P_{ab,j}
\right\rangle_t.
\label{eq:appC_CN_operator}
\end{equation}
This expression is linear in the density matrix and must not be replaced by an
expectation involving only the phase gradients.

Because $D_{ab}$ commutes with $g_{ij}(r_{ab})$, integration by parts gives the
polar representation
\begin{equation}
\begin{aligned}
\mathcal C_N
={}&
\frac{\gamma\kappa}{m_0^2}
\sum_{a<b}
\int dX_N\,\rho_N g_{ij}(r_{ab})
\\
&\times\Big[
D_{ab,i}S_ND_{ab,j}S_N
+
D_{ab,i}\ln\sqrt{\rho_N}
\\
&\hspace{19mm}\times
D_{ab,j}\ln\sqrt{\rho_N}
\Big].
\end{aligned}
\label{eq:appC_CN_polar}
\end{equation}
The second term is the amplitude contribution omitted if
$\widehat{\boldsymbol P}_{ab}$ is replaced by the gradient of the phase.

\subsection{Dilute and compact limits}

For the Gaussian kernel,
\begin{equation}
g_{ij}(r)
=
\left(
\frac{r_ir_j}{4r_C^4}
-
\frac{\delta_{ij}}{2r_C^2}
\right)g(r).
\label{eq:appC_gij_gaussian}
\end{equation}
The radial eigenvalue changes sign at $r=\sqrt{2}\,r_C$, so the pair term is not
sign-definite at arbitrary separation. In the dilute regime, $r_{ab}\gg r_C$, it
is exponentially suppressed and the balance reduces to the sum of one-particle
contributions.

In the compact regime, $r_{ab}\ll r_C$,
\begin{equation}
g_{ij}(r_{ab})
\simeq
-\frac{g(0)}{2r_C^2}\delta_{ij},
\end{equation}
so that
\begin{equation}
\mathcal C_N
\simeq
-\mu_C
\sum_{a<b}
\left\langle
\widehat{\boldsymbol P}_{ab}^{\,2}
\right\rangle_t,
\qquad
\mu_C
=
\frac{\gamma\kappa g(0)}{2m_0^2r_C^2}
=
\frac{\gamma\kappa C_d}{m_0^2}.
\label{eq:appC_CN_compact}
\end{equation}
The operator identity
\begin{equation}
\sum_{a<b}
\widehat{\boldsymbol P}_{ab}^{\,2}
=
(N-2)
\sum_{a=1}^N\widehat{\boldsymbol p}_a^{\,2}
+
\widehat{\boldsymbol P}_{\rm tot}^{\,2},
\qquad
\widehat{\boldsymbol P}_{\rm tot}
=
\sum_{a=1}^N\widehat{\boldsymbol p}_a,
\label{eq:appC_pair_operator_identity}
\end{equation}
then gives
\begin{equation}
\begin{aligned}
\frac{d}{dt}\mathbb E[\mathcal E_N]
={}&
NP_{\rm eff}^{(1)}
-
\left[
\Gamma_{\rm eff}^{(1)}+2m\mu_C(N-2)
\right]
\mathbb E[\mathcal E_N]
\\
&-
\mu_C\,
\mathbb E\!\left[
\left\langle
\widehat{\boldsymbol P}_{\rm tot}^{\,2}
\right\rangle_t
\right]
\end{aligned}
\label{eq:appC_compact_balance_Ptot}
\end{equation}
to leading compact order. A vanishing mean total momentum does not remove the last
term; the total-momentum variance must also be negligible.

For $N>2$, if
\begin{equation}
\mathbb E\!\left[
\left\langle
\widehat{\boldsymbol P}_{\rm tot}^{\,2}
\right\rangle_t
\right]
\ll
2m(N-2)\mathbb E[\mathcal E_N],
\label{eq:appC_Ptot_closure_condition}
\end{equation}
the leading compact balance closes as
\begin{equation}
\frac{d}{dt}\mathbb E[\mathcal E_N]
=
NP_{\rm eff}^{(1)}
-
\Gamma_N^{\rm comp}\mathbb E[\mathcal E_N],
\label{eq:appC_compact_closed_balance}
\end{equation}
where
\begin{equation}
\Gamma_N^{\rm comp}
=
\frac{\gamma\kappa m C_d}{m_0^2}
\left[
2N-
\frac{3(d+2)}{2}\chi
\right].
\label{eq:appC_Gamma_compact}
\end{equation}
The one-particle damping condition in Eq.~\eqref{eq:appC_positive_damping} is
sufficient to make Eq.~\eqref{eq:appC_Gamma_compact} positive for every $N>2$.
The stationary compact energy is
\begin{equation}
\mathcal E_{N,{\rm eq}}^{\rm comp}
=
\frac{NP_{\rm eff}^{(1)}}{\Gamma_N^{\rm comp}}
=
N\mathcal E_{1,{\rm eq}}
\frac{\Gamma_{\rm eff}^{(1)}}{\Gamma_N^{\rm comp}}.
\label{eq:appC_EN_compact_eq}
\end{equation}
Defining
$\mathcal E_{N,{\rm eq}}^{\rm comp}=(dN/2)k_BT_N^{\rm comp}$ gives
\begin{equation}
T_N^{\rm comp}
=
T_{\rm eff}
\frac{
1-\dfrac{3(d+2)}{8}\chi
}{
\dfrac{N}{2}-\dfrac{3(d+2)}{8}\chi
}.
\label{eq:appC_TN_compact}
\end{equation}
This quantity is only a parametrization of the closed leading-order energy balance.
For $N=2$, the internal-energy term in
Eq.~\eqref{eq:appC_pair_operator_identity} vanishes, and the compact correction is
controlled entirely by the total-momentum variance. Beyond the compact approximation,
Eq.~\eqref{eq:appC_CN_operator} depends on spatial and momentum correlations and does
not close in terms of the mean kinetic energy alone.



\begin{thebibliography}{99}

\bibitem{sakurai}
J. J. Sakurai and J. Napolitano,
\textit{Modern Quantum Mechanics}, 3rd ed.
(Cambridge University Press, Cambridge, 2020).

\bibitem{bohr1}
N. Bohr,
Nature (London) \textbf{121}, 580 (1928).

\bibitem{bohm1}
D. Bohm,
Phys. Rev. \textbf{85}, 166 (1952).

\bibitem{bohm2}
D. Bohm and B. J. Hiley,
\textit{The Undivided Universe: An Ontological Interpretation of Quantum Theory}
(Routledge, London, 2006).

\bibitem{Bassi2013}
A. Bassi, K. Lochan, S. Satin, T. P. Singh, and H. Ulbricht,
Rev. Mod. Phys. \textbf{85}, 471 (2013).

\bibitem{GRWref}
G. C. Ghirardi, A. Rimini, and T. Weber,
Phys. Rev. D \textbf{34}, 470 (1986).

\bibitem{DPref1}
L. Di\'osi,
Phys. Rev. A \textbf{40}, 1165 (1989).

\bibitem{DPref2}
R. Penrose,
Gen. Relativ. Gravit. \textbf{28}, 581 (1996).

\bibitem{CSLref}
G. C. Ghirardi, P. Pearle, and A. Rimini,
Phys. Rev. A \textbf{42}, 78 (1990).

\bibitem{Adlerref}
S. L. Adler,
J. Phys. A: Math. Theor. \textbf{40}, 2935 (2007).

\bibitem{NimmrichterHornberger2013}
S. Nimmrichter and K. Hornberger,
Phys. Rev. Lett. \textbf{110}, 160403 (2013).

\bibitem{lindref1}
V. Gorini, A. Kossakowski, and E. C. G. Sudarshan,
J. Math. Phys. \textbf{17}, 821 (1976).

\bibitem{lindref2}
G. Lindblad,
Commun. Math. Phys. \textbf{48}, 119 (1976).

\bibitem{holland95}
P. R. Holland,
\textit{The Quantum Theory of Motion: An Account of the de Broglie--Bohm
Causal Interpretation of Quantum Mechanics}
(Cambridge University Press, Cambridge, England, 1993).

\bibitem{kiefer92}
C. Kiefer,
Phys. Rev. D \textbf{45}, 2044 (1992).

\bibitem{ivanov20}
M. G. Ivanov, A. E. Kalugin, A. A. Ogarkova, and S. L. Ogarkov,
Symmetry \textbf{12}, 1657 (2020).

\bibitem{sint94}
S. Sint,
Nucl. Phys. B \textbf{421}, 135 (1994).

\bibitem{kieferwipf94}
C. Kiefer and A. Wipf,
Ann. Phys. (N.Y.) \textbf{236}, 241 (1994).

\bibitem{Smirne2014}
A. Smirne, B. Vacchini, and A. Bassi,
Phys. Rev. A \textbf{90}, 062135 (2014).

\bibitem{Smirne2015}
A. Smirne and A. Bassi,
Sci. Rep. \textbf{5}, 12518 (2015).

\bibitem{Toros2017}
M. Toro\v{s}, G. Gasbarri, and A. Bassi,
Phys. Lett. A \textbf{381}, 3921 (2017).

\bibitem{DiBartolomeo2023}
G. Di Bartolomeo, M. Carlesso, K. Piscicchia, C. Curceanu,
M. Derakhshani, and L. Di\'osi,
Phys. Rev. A \textbf{108}, 012202 (2023).

\bibitem{DiBartolomeo2024}
G. Di Bartolomeo and M. Carlesso,
New J. Phys. \textbf{26}, 043006 (2024).

\bibitem{Melo2026}
P. B. Melo, P. V. Paraguass\'u, S. Artini, G. Lo Monaco,
S. Donadi, and M. Paternostro,
arXiv:2606.06259 (2026).

\end{thebibliography}
\end{document}